%% file: osak_arxiv.tex
\documentclass[notitlepage,a4paper,pre,twocolumn,longbibliography]{revtex4-1}
\usepackage[utf8x]{inputenc}
\usepackage{inputenc}
\usepackage{graphicx}  
\usepackage{epstopdf}
\usepackage{bm}        
\usepackage{amsmath}   
\usepackage{amssymb}   
\usepackage{amsmath}
\usepackage{amssymb}
\usepackage{url}
\usepackage{siunitx} 
\usepackage{upgreek}
\usepackage{longtable}
\usepackage{array}
\usepackage[retainorgcmds]{IEEEtrantools}
\usepackage{MLHMM}
\renewcommand{\Eq}[1]{Eq.~\eqref{#1}} 

\usepackage{xr}
\usepackage{color}

\newcommand{\vc}[1]{\textcolor{blue}{#1}}

\date{\today}
\begin{document}
\title{Certain uncertainty: using pointwise error estimates in
  super-resolution microscopy}
\author{Martin Lindén\normalfont\textsuperscript{§,}}
\author{Vladimir \'{C}uri\'{c}\normalfont\textsuperscript{§,}}
\author{Elias Amselem}

\author{Johan Elf}
\email[]{johan.elf@icm.uu.se}
\affiliation{Department of Cell and Molecular Biology, Uppsala University, Sweden.\normalfont\textsuperscript{§} Contributed equally.}
\begin{abstract}
\input{abstract.tex}

\end{abstract}

\maketitle

\input{maintext.tex}

\section*{Methods}
\input{methods.tex}

\subsection*{Acknowledgements}
\input{acknowledgements.tex}
\subsection*{Author contributions}
\input{contributions.tex}
\input{osak_arxiv.bbl}
\clearpage
\onecolumngrid
\appendix
\section*{Supplementary material}

\renewcommand{\thesection}{S\arabic{section}}  
\setcounter{section}{0}    
\renewcommand{\thesubsection}{S\arabic{section}.\arabic{subsection}}  
\renewcommand{\thetable}{S\arabic{table}}  
\renewcommand{\thefigure}{S\arabic{figure}}
\setcounter{figure}{0}    
\renewcommand{\theequation}{S\arabic{equation}}
\setcounter{equation}{0}    

\input{supplementarytext.tex}

\end{document}

%% file: abstract.tex
Point-wise localization of individual fluorophores is a critical step
in super-resolution microscopy and single particle tracking. Although
the methods are limited by the accuracy in localizing individual
flourophores, this point-wise accuracy has so far only been estimated
by theoretical best case approximations, disregarding for example
motional blur, out of focus broadening of the point spread function
and time varying changes in the fluorescence background. Here, we show
that pointwise localization uncertainty can be accurately estimated
directly from imaging data using a Laplace approximation constrained
by simple mircoscope properties.  We further demonstrate that the
estimated localization uncertainty can be used to improve downstream
quantitative analysis, such as estimation of diffusion constants and
detection of changes in molecular motion patterns.  Most importantly,
the accuracy of actual point localizations in live cell
super-resolution microscopy can be improved beyond the information
theoretic lower bound for localization errors in individual images, by
modeling the fluorophores' movement and accounting for their
point-wise localization uncertainty.

%% file: maintext.tex
\begin{figure}[b]
\includegraphics[width=80mm]{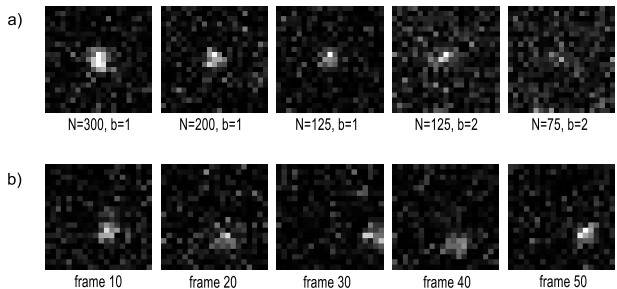}
\caption{\label{fig:frameExamples}
Simulated images of a diffusing fluorophore with varying localization
accuracy.  (a) Fluorescent spots under different imaging conditions,
where $N$ is the number of photons per spot and $b$ is the background
level of photons per pixel; (b) Representative frames from a simulated
movie with $b=1,N=150$. Images were simulated using
SMeagol \cite{linden2016}.}
\end{figure}

Super-Resolution fluorescence microscopy and live cell single particle
tracking rely on computer intensive data analysis to find and localize
single fluorescent emitters in noisy images. Much effort has been
spent on developing and testing efficient spot localization algorithms
\cite{sage2015} and understanding the theoretical limits for
localization accuracy
\cite{thompson2002,ober2004,mortensen2010,rieger2014}. However, the
problem of estimating and using the actual accuracy is still unsolved.

PALM/STORM type super-resolution imaging \citep{betzig2006, rust2006}
relies on the serial activation and localization of sparse
photo-switchable fluorophores. Knowledge about the localization
accuracy is important to build up a high resolution image since
uncertain points will only contribute blur. Often, only the number of
photons, pixel size, and background noise for each emitter is used to
estimate the localization accuracy, assuming that it achieves its
theoretical limit. However, theoretical estimates neglect many
important factors, and are prone to systematic errors in particular
when the background is variable and the emitter is moving, which is
the common situation for live cell super-resolution imaging.

Knowledge of the localization uncertainty is also important in single
particle tracking (SPT) \cite{manley2008}, where it can be used to
improve estimators of diffusion constants
\cite{vestergaard2014,relich2016}.  It is common in live cell imaging
that the localization uncertainty varies throughout an experiment, for
example due to out-of-focus motion, drift, motion blur, fluorophore
intensity fluctuations, heterogeneous background, or gradual
photobleaching of the background or labeled molecule. Examples of
heterogeneous and time-varying spot quality are shown in
Fig.~\ref{fig:frameExamples}.

Here, we investigate methods to extract and use localization
uncertainty of single dots in super-resolved single particle tracking,
using a combination of experimental data and highly realistic
simulated microscopy experiments \cite{linden2016}.  We propose and
characterize an uncertainty estimator based on the Laplace
approximation, combined with information about physical limitations in
the detection system.
This method outperforms the common practice of combining
maximum-likelihood localization with a Gaussian point-spread function
(PSF) model, and the Cram\'er-Rao lower bound (CRLB), which
systematically underestimates the uncertainty in low light conditions
relevant for live cell applications.

Second, we demonstrate how estimated localization uncertainties can be
used to improve the estimation of diffusion constants, particle
positions, and state changes in single- and multi-state diffusion SPT
data.  
For the multi-state case, we derive a variational Expectation
Maximization (EM) algorithm for a diffusive Hidden Markov model (HMM)
which extends previously described algorithms
\cite{calderon2016,relich2016,bernstein2016,koo2015,slator2015,calderon2014,vestergaard2014}
by accounting for both multi-state diffusion, localization
uncertainty, and motion blur.



\section*{Results}
\begin{figure}
\includegraphics[width=80mm]{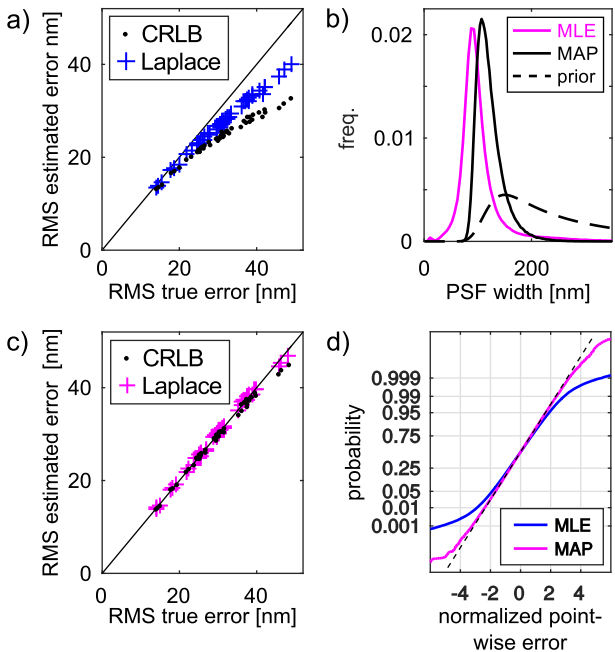}
\caption{\label{fig:simErr} Actual and estimated errors in simulated
  data.  (a) Root mean square (RMS) error
  $\sqrt{\mean{(\mu_\text{est.}-\mu_\text{true})^2}}$ vs.~root mean
  estimated error variance $\sqrt{\mean{\varepsilon^2}}$ for MLE
  localization from the CRLB and Laplace approximations.  (b) PSF
  width prior and fitted parameter distributions with (MAP) and
  without (MLE) prior. (c) Actual and estimated errors for MAP
  localization. (d) Probability plot of point-wise errors normalized
  by point-wise Laplace uncertainty estimates,
  $(\mu_\text{est.}-\mu_\text{true})/\sqrt{\varepsilon^2}$. The dashed
  line indicates the reference standard Gaussian distribution.}
\end{figure}

\subsection*{Estimating point-wise uncertainty}
Estimating uncertainty is closely related to estimating positions,
where the maximum likelihood estimate (MLE) is generally considered
the optimal method. A maximum likelihood method starts with a
likelihood function, i.e., the probability density function of a
probabilistic model for generating images of spots (pixel counts in a
small region around a spot) with the emitter position among the
adjustable parameters. The MLE are the parameters that maximize the
likelihood function for a particular spot image.  Following common
practice, we model EMCCD camera noise with the high-gain approximation
and Gaussian readout noise \cite{chao2013,mortensen2010}, and the spot
shape by a symmetric Gaussian intensity profile plus a constant
background intensity. The fit parameters are thus spot position
$(\mu_x,\mu_y)$, background intensity $b$, PSF width $\sigma$, and
spot amplitude $N$ (see Methods, \Eq{eq:gaussPSF}), while the camera
noise is assumed known from calibration.

What is localization uncertainty?  The localization error is the difference
$\mu_\text{est.}-\mu_\text{true}$ between estimated and true
positions. By localization uncertainty, we seek the distribution of
the error, either in a Bayesian posterior sense, or in the sense of
repeated localizations of spots with the same uncertainty.  The
uncertainty of the localization is related to the shape of the
likelihood maximum: a sharply peaked maximum means that only a narrow
set of parameters are likely, while a shallow maximum means greater
uncertainty.  

The Cram\'er-Rao lower bound (CRLB) is the smallest possible
  variance of an unbiased estimator for a given set of model
  parameters, and is related to the expected sharpness of the
  likelihood maximum (see Methods, \Eq{eq:CRLB}).
While this is strictly speaking not a statement about a single image,
but rather about the average information content of data generated by
a model, it is often used to estimate localization uncertainty. A
Bayesian alternative is the Laplace approximation \cite{mackay2003},
which derives an approximate Gaussian posterior distribution for the
fit parameters in terms of the sharpness of the likelihood maximum for
each particular image (see Methods, \Eq{eq:laplace}).  Both estimators
supply an estimated variance $\varepsilon^2$, but none of them are
well characterized as estimators of localization uncertainty.

To test these estimators, we analyzed a large set of simulated movies
of a fluorescent particle diffusing at D=1
\si{\square\micro\metre\per\second} in an \textit{E. coli}-like
geometry. The movies cover a broad range of experimentally relevant
imaging conditions (see Methods) and include realistic EMCCD noise,
background fluorescence, a non-Gaussian and space-dependent PSF
\cite{kirshner2013}, and motion blur \cite{deschout2012}.

A basic consistency check is that the average estimated error
variance, $\mean{\varepsilon^2}$, agrees with the variance of the
actual errors, $\mean{(\mu_\text{est.}-\mu_\text{true})^2}$.  In
Fig.~\ref{fig:simErr}a, we compare the square root of these quantities
for different imaging conditions, based on MLE localizations combined
with either CRLB or Laplace uncertainty estimators. We obtain
consistency under good imaging conditions, where the spots are bright
and the average errors low. However, as conditions worsen and the
errors increase, the uncertainty is underestimated, especially by the
CRLB.

There are several possible reasons for this behavior. The Laplace
approximation is based on a truncated Taylor expansion of the log
likelihood (see Methods). The CRLB is strictly not a point-wise error
estimator at all, and is further defined in terms of the true
parameter values, although by necessity evaluated with the fitted
ones. The simplified Gaussian PSF model performs well for localization
\cite{mortensen2010}, but this does not guarantee good uncertainty
estimates. Any of these approximations might fail in noisy or low
light conditions.  Another possibility is that localizations under
poor imaging conditions are corrupted by a sub-population of fits that
converge to a local minimum that does not reflect the underlying spot
shape, e.g., fitting a single bright pixel as a very narrow spot, or
misinterpreting the PSF shoulders as background or vice versa.

To address the latter complication, we constrained the localizations
using prior distributions on selected parameters, thus replacing MLE
with maximum a aposteriori estimation (MAP). Fixing parameter values
is not suitable here, since both the size and shape of spots fluctuate
due to fluorophore motion and varying imaging conditions. Furthermore,
it is experimentally easier to obtain independent information about
the background and PSF width than about the spot intensity. We
therefore limit our attention to background and PSF width, and found
the following priors to perform well:
a log-normal prior centered on the true value for the background
intensity, and a skewed log-normal \cite{azzalini1985} prior for the
PSF width to penalize fits with unphysical widths below that of an
in-focus spot.

The PSF width prior together with the distribution of fitted PSF
widths are shown in Fig.~\ref{fig:simErr}b. PSF widths below 1 pixel
(80 \si{\nano\metre}) are virtually eliminated in the MAP
fits. Background and spot amplitudes (see Fig.~\ref{SI:fig:parPrior})
are shifted somewhat downwards and upwards, respectively.  As seen in
Fig.~\ref{fig:simErr}c, this substantially improves the agreement
between true and estimated errors under all imaging conditions.  The
Laplace estimator still outperforms the CRLB by a small margin, and
numerical experiments on a wider range of priors (see SI
Fig.~\ref{SI:fig:priors_SI}) further confirm that the Laplace
estimator is more robust to non-optimal priors than the CRLB.

How does the prior help?  A direct comparison of the true errors for
the MLE and MAP fits (SI Fig.~\ref{SI:fig:MLE_vs_MAPerrors}) reveals
very small differences, indicating that the improvement is mainly due
to improved uncertainty estimates.  This is consistent with the
theoretical observation that the Fisher information matrix for the
localization problem is nearly block-diagonal \citep{mortensen2010}
with very weak coupling between two groups of fitting parameters:
positions ($\mu_x,\mu_y$) and shape parameters ($N,b,\sigma$).  Thus,
additional information about one of these groups does not reduce the
errors of the other significantly.  For uncertainty estimation, this
is not the case: information about the background and PSF shape does
help in estimation position uncertainty.

While mean-square errors are useful, we are ultimately interested in
the full distribution of errors. In particular, most
\cite{berglund2010,calderon2016,relich2016,bernstein2016,koo2015,slator2015,calderon2014,vestergaard2014}
(but not all \cite{ashley2015}) statistical models of SPT data assume
Gaussian errors, but this assumption has not been tested.  The Laplace
approximation (\Eq{eq:laplace}) is a Gaussian approximation. If it was
exact, the errors normalized by the estimated standard deviation would
be Gaussian with unit variance, and produce a straight line in the
Gaussian probability plot of Fig.~\ref{fig:simErr}d.  The actual
normalized errors based on MAP localizations is more Gaussian than the
MLE ones, and follow the straight line for almost four standard
deviations. This shows that the improved consistency of the MAP
estimates also translates to more Gaussian error distributions.

Overall, these results show that point-wise uncertainty estimation
using the Laplace approximation works well in a wide range of
experimentally relevant conditions, but that experimentally accessible
additional information about background and PSF shape is necessary to
get consistent results in low light conditions. Moreover, we see good
support for the common assumption of Gaussian errors.

\begin{figure*}
\centering
\includegraphics[width=160mm]{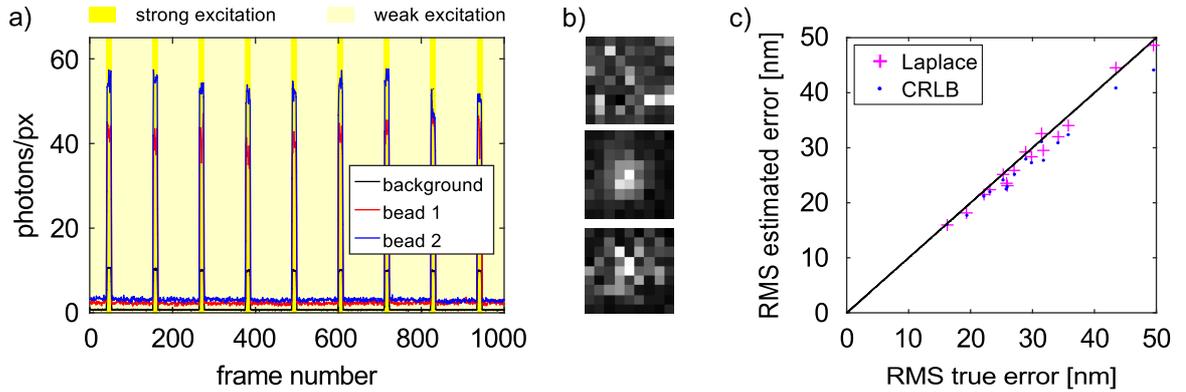}
\caption{\label{fig:realData} Validating estimators of localization
  uncertainty using real data.  (a) Intensity in different frames for
  two different beads and the background. Excitation intensity
  indicated by yellow background. (b) Image examples. Top: background
  in high intensity frames. Middle: Bead 1 in a high intensity
  frame. Bottom: Bead 1 in a low intensity frame.  (c) Actual and
  estimated RMS errors for different beads.}
\end{figure*}

\subsection*{Validation on real data}
To test the above conclusions on real data, we imaged immobilized
fluorescent beads, alternating strong and weak excitation as shown in
Fig. \ref{fig:realData}a. We used images under strong excitation
conditions to extract a high accuracy ground truth for testing the
uncertainty estimates in the dim images.  We estimated the position
and uncertainty of spots using the MAP estimation described above,
with a background prior centered around the mean background (0.8
photons/pixel) seen in dim frames. A drift-corrected ground truth was
estimated by cubic spline interpolation between the mean positions
obtained from each block of 10 consecutive bright images.  Since 
the intensity differs by about a factor 10 between bright and dim
frames, the RMS errors of the ground truth should be approximately
10-fold lower than that of a single dim spot. Fig.~\ref{fig:realData}c
shows the resulting error-uncertainty comparison, with every point
corresponding to a single bead. It reproduces the behavior on
simulated images in Fig.~\ref{fig:simErr}b, thus confirming our
conclusion that the Laplace approximation is preferable to CRLB for
uncertainty estimators, and that the good performance of our proposed
PSF width prior is not limited to that particular simulated data set.

\begin{figure}[t]
\centering
\includegraphics[width=80mm]{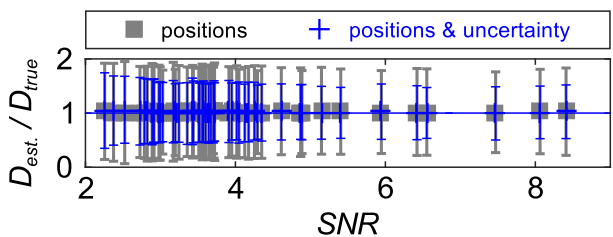}
\caption{\label{fig:Dest} Estimated diffusion constants
  vs.~signal-to-noise ratio $SNR=\sqrt{D\Delta
    t/\mean{\varepsilon^2}}$ \cite{vestergaard2014} in simulated
  10-step trajectories, mean value and 1\% quantiles.}
\end{figure}

\subsection*{Diffusion constants}

Next, we consider how estimated localization uncertainties leads to
better estimates of diffusion constants, arguably the most common
analysis of single particle tracking data. We divided the synthetic
data shown in Fig.~\ref{fig:frameExamples} into trajectories of length
10, estimated positions and uncertainties with the MAP and Laplace
estimators described above, and estimated diffusion constants using
the covariance-based estimators of Ref.~\cite{vestergaard2014} with
and without the use of uncertainty estimates.  Fig.~\ref{fig:Dest}
shows the mean value and 1\% quantiles of the two estimators applied
to every imaging condition, plotted against the signal-to-noise ratio.
As expected, the use of estimated uncertainties improves the
variability of the diffusion estimates substantially. The
covariance-based estimators only use the average uncertainty in each
trajectory (see Methods). We also implemented a maximum likelihood
estimator for the diffusion constant \cite{relich2016} which makes
explicit use of the point-wise uncertainties (see SI text
\ref{SI:sec:diffusionMLE}), but found no further improvement.

\begin{figure}
\includegraphics[width=80mm]{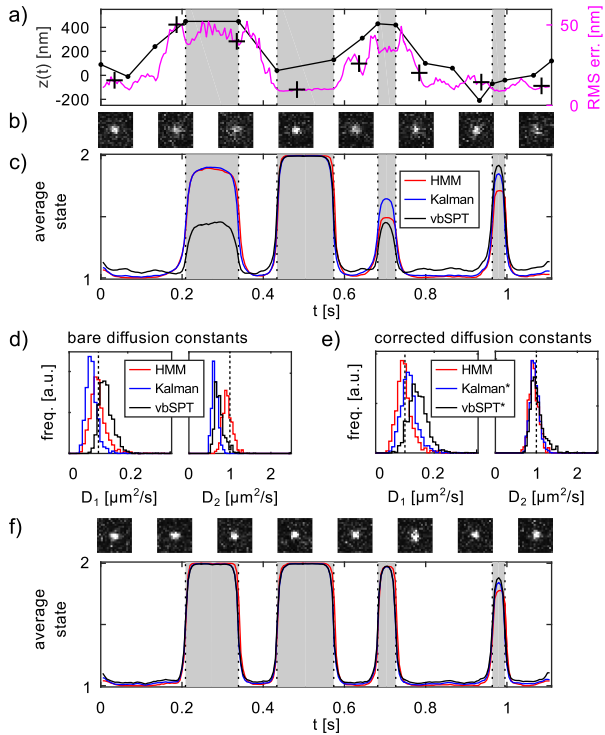}
\caption{\label{fig:HMM} Detecting binding events in and out of focus.
  (a) Input $z$-coordinates and RMS errors, and (b) representative
  spot images along simulated trajectories, from time points indicated
  by + in (a). Gray areas indicate binding events. (d) Average state
  occupancy for the different analysis algorithms ( ``HMM n.b.'' =
  no-blur HMM).  (e) Distribution of estimated diffusion constants for
  the two states, with true values indicated by dashed lines. (f)
  Corrected diffusion constant estimates. (g) Spot images and HMM
  occupancies for trajectories simulated without defocus effects.}
\end{figure}

\subsection*{Analysis of multi-state data}
We now turn to a more challenging problem where point-wise errors do
matter: data where both the diffusion constant and localization error
change significantly on similar time scales. In single particle
tracking, changes in diffusion constant can be used as a non-invasive
reporter on intracellular binding and unbinding events
\cite{persson2013}. However, diffusive motion and localization errors
contribute additively to the observed step length statistics (see
Eq.~\ref{MT:eq:dxcov}), and thus changes in diffusion constants and
localization errors cannot be reliable distinguished.

As an example, we consider a protein that alternates between free
diffusion ($D=1$ \si{\square\micro\metre\per\second}) and a bound
state simulated by slow diffusion ($D=0.1$
\si{\square\micro\metre\per\second}).  We focus on a single trajectory
with 4 binding/unbinding events, two of which occur about 400 nm out
of focus, and thus are accompanied by substantial broadening of the
PSF and accompanying increases in localization errors. This defocus
matches roughly the radius of an \textit{E.~coli} cell, and the
scenario could model tracking experiments with cytoplasmic proteins
that can bind to the inner cell membrane.

Using SMeagol \cite{linden2016}, we simulated 12 000 replicas of the
above set of events, at a camera frame rate of 200 \si{\hertz},
continuous illumination, and 300 photons/spot on
average. Fig.~\ref{fig:HMM}a shows the $z$ coordinates in the input
trajectory, and the frame-wise RMS errors produced by the MAP
localization algorithm described above. The input points are sparse
(tens of \si{\milli\second} apart), which allows SMeagol to produce
simulated movies that contain the same binding events and defocus
trends, but vary in the detailed diffusion trajectory as well as in
noise realizations. Examples of simulated spots along a trajectory are
shown in Fig.~\ref{fig:HMM}b.

To analyze this challenging data set, we extended the maximum
likelihood diffusion constant estimator with explicit point-wise
errors to include multiple diffusion states governed by an hidden
Markov model (HMM), for which we derived a variational EM algorithm
(see SI text \ref{SI:sec:HMM}). We then analyzed each simulated
trajectory with three different 2-state HMMs: (i) vbSPT, which models
the observed positions as pure diffusion and neglects blur and
localization errors \cite{persson2013}, (ii) the above-mention HMM
that explicitly models these effects, and (iii) the Kalman filter
limit of the HMM in (ii), which models localization errors but not
blur effects. Since multi-state Kalman-type algorithms have been
studied previously \cite{bernstein2016,slator2015,calderon2014}, it is
interesting to compare models with and without blur.
Fig.~\ref{fig:HMM}c shows the inferred average state from the three
different models. As expected, the HMMs that include localization
errors outperform vbSPT at detecting the strongly defocused first and
third binding events. The full HMM does not give the best
classification of the two short binding events, but it does give the
lowest overall misclassification rate, 9\% versus 9.6\% and 17\% for
the Kalman and vbSPT models, respectively, so the apparent worse time
resolution might reflect an overall tendency to invent spurious
transitions by the Kalman and vbSPT models.

Next, we look at estimated diffusion constants. Here, the Kalman and
vbSPT models make systematic errors as seen in the bare parameters in
Fig.~\ref{fig:HMM}d. However, by comparing the step length statistics
between the full and simplified models, one can derive heuristic
correction factors for these models (see Methods, \Eq{eq:Dcorr}). As
shown in Fig.~\ref{fig:HMM}e, this reduces the bias substantially,
especially for the high diffusion constant.

To finally compare the different HMMs on more well-behaved data, we
reran the same experiment but with all $z$ coordinates rescaled by a
factor 1/5 in the PSF model, which essentially removes the
$z$-dependent defocus effects. On this less challenging data set,
event detection is much improved and the differences between the three
HMMs are much less pronounced. However, with misclassification rates
of 4.8\%, 6.6\%, and 7.6\% for the HMM, Kalman model, and vbSPT,
respectively, the full HMM still does overall better.

\subsection*{Position refinement}
\begin{figure}
\includegraphics[width=80mm]{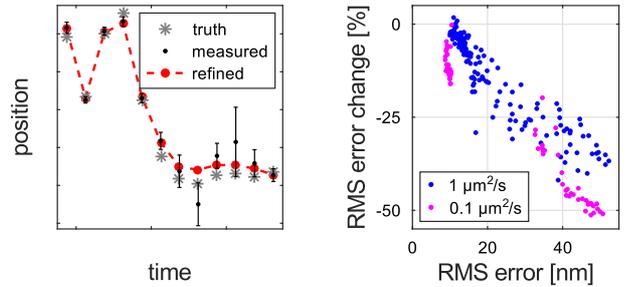}
\caption{\label{fig:pos} 
Improved localization accuracy by modeling particle motion.  (a) True,
measured, and HMM-refined positions from part of a 2-state
trajectory. (b) Relative change of RMS localization error after
HMM-based refinement, for every frame in Fig.~\ref{fig:HMM}a.}
\end{figure}

Since the new HMM includes the true trajectory as a hidden variable
and performs a global analysis, it can be used to refine individual
localized positions, and in principle beat the Cramer-Rao lower bound
for single image localizations. Fig.~\ref{fig:pos}a shows the true,
measured, and refined positions for part of a two-state trajectory,
and Fig.~\ref{fig:pos}b the relative change of the RMS error for each
frame in Fig.~\ref{fig:HMM}a after refinement. The refinement improves
the RMS errors with up to 50\%. Large localization errors and small
diffusion constant leads to larger relative improvement, which is
intuitive since those factors both make a single point less
informative relative its neighbors.  A few points show a small error
increase. On closer inspection, these turn out to be situated near
hard-to-detect state-changes, and therefore tend to be refined using an
incorrect diffusion constant.

\section*{Discussion}
Fluorophore positions are not the only useful kind of information in
super-resolution microscopy images.  Here, we have shown that
point-wise position uncertainty can also be extracted and used to
improve quantitative data analysis.  This is particularly important
for live cell data where dynamic phenomena can be studied, and one may
expect more heterogeneous imaging conditions where many of the
theoretical estimates of localization errors, that may apply for cells
with immobilized internal structure (``fixed''), will have little
relevance.

In general, our results indicate that estimation of position
uncertainty is more sensitive to the fit model than estimation of
position. For parctical use, we find that an estimate based on the
Laplace approximation combined with external information about the
fluorescent background and PSF shape performs well in a wide range of
experimentally relevant conditions.

An intuitive reason why the CRLB performs worse may be that it makes
explicit use of the fit model twice, for both fitting and predicting
the model-dependent uncertainty, while the Laplace approximation only
requires the first step. 
Theoretically, the posterior density formalism of the Laplace
approximation also fits more naturally with model-based time-series
analysis where the particle trajectory is treated as a latent variable
to be integrated out.

While we have focused on 2D localization using conventional optics,
the extension to e.g., three dimensions using dual plane imaging
\cite{liu2013} or engineered PSFs \cite{pavani2009} present no
principal difficulties, as long as appropriate PSF models for
localization can be formulated. Note however that even a perfect
characterization of an experimental PSF \cite{liu2013} does not
constitute a perfect localization model, since it does not describe
random PSF shape fluctuations due to motion blur
\cite{deschout2012}. Current techniques for 3D localization are
inherently asymmetric and yield different in-plane and axial accuracy
\cite{rieger2014}, which further underscores the need for downstream
analysis methods to incorporate heterogeneous localization
uncertainty.

Most super-resolution microscopy applications are however not aimed at
particle tracking, but imaging. For PALM/STORM type imaging of fixed
samples, the ability to estimate the uncertainty of individual spots
does not improve the localizations themselves, but it may still
improve the final resolution since uncertain points can be omitted on
a quantitative basis. However, the consequences for live cell imaging
are more interesting, since the same fluorophore may be detected in
different positions over different frames if the target is moving. For
this case, we have shown that the combination of estimating
uncertainty and modeling the fluorophore motion can produce refined
position estimates, in principle pushing the localization errors below
the single-spot Cramer Rao lower bound, by merging information from
consecutive frames in an optimal way.

\nocite{chow2009,ghahramani1997,bishop2006,rabiner1986,bronson2009,johnson2014,burnham}


%% file: methods.tex
\subsection*{Synthetic data}
We generated synthetic microscopy data using SMeagol, a software for
accurate simulations of dynamic fluorescence microscopy at the single
molecule level \cite{linden2016}. We modeled the optics using a
Gibson-Lanni PSF model with 584 nm wavelength and NA=1.4
\cite{kirshner2013}, and an EMCCD camera with the high gain
approximation \cite{chao2013,mortensen2010} plus Gaussian readout
noise.

For localization and diffusion estimation tests, we simulated simple
diffusion ($D=1$\si{\square\micro\metre\per\second}) in a cylinder of
length 14.4 \si{\micro\metre} and diameter 2 \si{\micro\metre},
similar to long \textit{E.~coli} cells, in order to avoid confinement
artifacts in the longitudinal direction. We generated 48 data sets
spanning a wide range of imaging conditions by varying frame duration
(8 \si{\milli\second} or 10 \si{\milli\second}, exposure time 2
\si{\milli\second}, 4 \si{\milli\second}, 6 \si{\milli\second}, and 3
\si{\milli\second}, 5 \si{\milli\second}, 8 \si{\milli\second}
respectively, background intensity (1 or 2 photons per pixel), average
spot brightness (75-300 photons/spot), EM gain 20 or 30, and readout
noise level 4 or 8. Each data set contained $10^4$ individual spots.

For the simulated multi-state data, we hand-modified a single SMeagol
input trajectory from a simulated 2-state model to contain 4 binding
events with different durations and z-coordinates as seen
Fig.~\ref{fig:HMM}a, and also thinned out the input trajectory to
create more variability in the particle paths between different
realizations. We then simulated many realizations from this input
trajectory, using the same PSF model as above, continuous illumination
with a sample time of 5 \si{\milli\second}, EMCCD gain 90, an average
spot intensity of 300 photons/spot, and a time-dependent background
that decays exponentially from 0.95 to 0.75 background photons per
pixel with a time-constant of 0.75 s.
\subsection*{Real data}
For estimating localization errors in the real imaging conditions, we
use immobilized fluorescent beads with the diameter of 0.1
\si{\micro\metre} (TetraSpeck Fluorescent Microspheres, ThermoFischer
T7284). The beads where diluted in ethanol and then placed on a
coverslip where we let them dry in before adding water as a mounting
medium.

Imaging was done with a Nikon Ti-E microscope which was configured for
EPI-illumination with a 514 \si{\nano\metre} excitation laser
(Coherent Genesis MX STM) together with matching filters (Semrock
dichroic mirror Di02-R514 with emission filter Chroma HQ545/50M-2P
70351).  Intensity modulation was made possible by an acousto-optic
tunable filter (AOTF) (AA Opto Electronics, AOTFnC) which was
triggered by a waveform generator (Tektronix, AFG3021B). The waveform
used was a sequence of square pulses, high for 200 \si{\milli\second}
and low for 1800 \si{\milli\second}. The two illumination intensities,
high and low, corresponds to 10.7 \si{\kilo\watt\per\square\centi\metre}
and 0.63 \si{\kilo\watt\per\square\centi\metre}, 
respectively.

Fluorescent beads where viewed through a 100x (CFI Apo TIRF 100x oil,
Na=1.49) objective with a 2X (Diagnostic instruments DD20NLT)
extension in front an Andor Ultra 897 EMCCD camera (Andor Technology
Ltd.).
This configuration puts the pixel size to 80 \si{\nano\metre} which is
the same pixel size set in the simulated data.  The data set
constituted of 1000 frames (Fig. \ref{fig:realData}a) with an exposure
time of 30 \si{\milli\second}. EMCCD noise characteristics (gain,
offset, readout noise) were determined by analyzing a “dark” movie
obtained with the shutter closed.

\subsection*{Localization} 

We perform maximum likelihood (MLE) localization using a high-gain
EMCCD noise model \cite{chao2013,mortensen2010}, which relates the
probability $q(c_i|E_i)$ of the offset-subtracted pixel count $c_i$
for a given pixel intensity $E_i$ (expected number of photons/frame)
in pixel $i$. For the intensity $E(x,y)$, we model the PSF with a
symmetric Gaussian,
\begin{equation}\label{eq:gaussPSF}
E(x,y)=\frac{b}{a^2}+\frac{N}{\sqrt{2\pi\sigma^2}}
\exp\Big(-\frac{(x-\mu_x)^2+(y-\mu_y)^2}{2\sigma^2}\Big),
\end{equation}
with pixel size $a$, background $b$ (expected number of
photons/pixel), spot width $\sigma$, amplitude $N$ (expected number of
photons/spot), and spot position $(\mu_x,\mu_y)$, and approximate the
pixel intensity
\begin{equation}
E_i=\int_\text{px. $i$} E(x,y)dxdy
\end{equation}
by numerical quadrature \cite{chao2015}. The log likelihood of an
image containing a single spot is then given by
\begin{equation}
\ln L(\theta) = \ln q_0(\theta)+\sum_{i\in ROI} \ln q(c_i|E_i(\theta)),
\end{equation}
where $\theta=(\mu_x,\mu_y,b,N,\sigma)$ are fit parameters, and $q_0$
is a prior distribution (we set $\ln q_0=0$ for MLE fitting).  To
avoid the complications of spot identification, we use known positions
to determine the ROI and initial guess for $(\mu_x,\mu_y)$. All
localizations are performed using a $9\times 9$ ROI. Fits that failed
to converge or estimated uncertainties larger than 2 pixels where
discarded.


\subsection*{Cramer-Rao lower bound} 

The CRLB is a lower bound on the variance of an unbiased estimator
\cite{cramer1945,rao1945}. 
We use an accurate approximation to the CRLB for a symmetric Gaussian
PSF from Ref.~\cite{rieger2014},
\begin{equation}\label{eq:CRLB}
\varepsilon^2_\text{CRLB}=2\frac{\sigma_a^2}{N}\left(
1+4\tau+\sqrt{\frac{2\tau}{1+4\tau}}
\right), 
\end{equation}
with $\tau=2\pi\sigma_a^2 b/(Na^2)$, $\sigma_a^2=\sigma^2+a^2/12$, and
the prefactor 2 accounts for EMCCD excess noise \cite{mortensen2010}.

\subsection*{Laplace approximation}
An alternative way to approximate the uncertainty of the fit
parameters is to Taylor expand the likelihood around the
maximum-likelihood parameters $\theta^*$ to second order, that is
\begin{equation}\label{eq:laplace}
\ln L(\theta)\approx \ln L(\theta^*)+
\frac{\partial \ln L}{\partial\theta}\Big|_{\theta^*}(\theta-\theta^*)
    +\frac12(\theta-\theta^*)^T
    \frac{\partial^2\ln L}{\partial \theta^2}\Big|_{\theta^*}
(\theta-\theta^*).
\end{equation}
The first-order term is zero, since $\theta^*$ is a local maximum.
This approximates the likelihood by a Gaussian with covariance matrix
given by the inverse Hessian, i.e., $\Sigma=[\partial^2\ln L/\partial
  \theta^2]^{-1}$. In a Bayesian setting, this expresses the
(approximate) posterior uncertainty about the fit parameters.  The
estimated uncertainties (posterior variances) are given by the
diagonal entries of the covariance matrix, e.g.,
$\varepsilon^2_\text{Lap.}(\mu_x)=\Sigma_{\mu_x,\mu_x}$.  We compute
the Hessian numerically using Matlab's built-in optimization routines,
and use the log of the scale parameters $b,N,\sigma$ for fitting,
since they are likely to have a more Gaussian-like posterior
\cite{mackay1998}.
\subsection*{Prior distributions}
The prior distributions for localization used in the main text are:
normal priors with mean value $\ln b_\text{true}$ and std.~0.2 for the
log background intensity, resulting in the log-normal prior $b\in\ln
N(\ln b_\text{true},0.2^2)$.  For the PSF width $\sigma$, we use a
skewed normal prior for $\ln \sigma$ to penalize fits with $\sigma$
below the minimum in-focus width of the PSF. The skew normal density
is given by \cite{azzalini1985}
\begin{equation}
  \label{eq:skewed_prior}
  f(x,x_0,w,\alpha) = 
  \frac{2}{w}
  \phi\left(\frac{x-x_0}{w}\right)\Phi\left(\alpha\frac{x-x_0}{w}\right),
\end{equation}
where $\phi$ is the probability distribution function and $\Phi$ is
the cumulative distribution function for the unit normal distribution
$N(0,1)$. We used $x_0=\log(1.5)$, $w=1$, and $\alpha = 5$, as
sketched in Fig.~\ref{fig:simErr}b, with the pixel width $a=80$
\si{\nano\metre} as the length unit. These priors are shown in SI
Fig.~\ref{SI:fig:parPrior}.


\subsection*{Covariance-based diffusion estimator}
If $x_k$ ($k=0,1,\ldots$) is the measured trajectory of a freely
diffusing particle with diffusion constant $D$, the widely used model
for camera-based tracking by Berglund \cite{berglund2010} predicts
that the measured step lengths $\Delta_k=x_{k+1}-x_k$ are zero-mean
Gaussian variables with covariances given by
\begin{equation}\label{eq:dxcov}
\mean{\Delta_k^2}=2D\Delta t(1-2R)+2\varepsilon^2,\quad
\mean{\Delta_k\Delta_{k\pm1}}=2D\Delta tR-\varepsilon^2,
\end{equation}
and uncorrelated otherwise.  Here, $0\le R\le 1/4$ is a blur
coefficient that depends on how the images are acquired (e.g., $R=1/6$
for continuous illumination), $\Delta t$ is the measurement time-step,
and $\varepsilon^2$ is the variance of the localization errors.

Substituting sample averages for $\mean{\Delta_k^2}$ and
$\mean{\Delta_k\Delta_{k+1}}$ and solving for $D$ yields a
covariance-based estimator (CVE) with good performance
\cite{vestergaard2014}. If $\varepsilon^2$ is known or can be
estimated independently, the first relation in \Eq{eq:dxcov} alone
yields a further improved estimate of $D$. As we argue in
Sec.~\ref{SI:sec:cve}, these estimators apply also for variable
localization errors if $\varepsilon^2$ is replaced by the average
$\mean{\varepsilon^2}$.

\subsection*{Maximum likelihood and multi-state diffusion}
The Berglund model \cite{berglund2010} can also be used directly for
maximum likelihood inference, which has the potential advantage that
point-wise errors can be modeled \cite{relich2016}. The basic
assumption is to model the observed positions $x_k$ as averages of the
true diffusive particle path $y(t)$ during the camera exposure, plus a
Gaussian localization error, i.e.,
\begin{equation}\label{eq:blurdef}
x_k=\int_{0}^{\dt}y(k\dt+t)f(t)dt+\varepsilon_k\xi_k,
\end{equation}
where $f(t)$ is the normalized shutter function \cite{berglund2010},
$\varepsilon_k$ is the localization uncertainty (standard deviation)
at time $k$, and $\xi_k$ are independent Gaussians random numbers with
unit variance.  Continuous illumination is described by a constant
shutter function, $f(t)=1/\dt$. The limit where blur effects are
neglected can be described by setting $f(t)$ to a delta function,
which corresponds to instantaneous position measurement. This reduces
\Eq{eq:blurdef} to a standard Kalman filter \cite{calderon2016}, and
leads to $R=0$ in \Eq{eq:dxcov}.

In SI text \ref{SI:sec:diffusionMLE}, we derive a maximum likelihood
estimator that learns both $D$ and $y(t)$. In SI
Sec.~\ref{SI:sec:HMM}, we extend the model to multi-state diffusion,
by letting the diffusion constant switch randomly between different
values corresponding to different hidden states in an HMM, and derive
a variational EM algorithm for maximum likelihood inference of model
parameters, hidden states, and refined estimates of the measured
positions.

To interpret estimated diffusion constants from simplified models, one
may ``derive'' corrected diffusion estimates $D^*$ by equating
expressions for the step length variance $\mean{\Delta_k^2}$ from
\Eq{eq:dxcov} with and without those effects present.  For the Kalman
($R=0$) and vbSPT ($R=\varepsilon=0$) models, we get
\begin{equation}\label{eq:Dcorr}
  D^*=\frac{D_\text{Kalman}}{1-2R},\quad \text{and }
  D^*=\frac{D_\text{vbSPT}-\mean{\varepsilon^2}/\dt}{1-2R},
\end{equation}
respectively, which is what we use in Fig.~\ref{fig:HMM}e.

\subsection*{Software}
Matlab open source code for the EM and localization algorithms are
available at
\href{https://github.com/bmelinden/uncertainSPT}{github.com/bmelinden/uncertainSPT}.

%% file: acknowledgements.tex
We thank David Fange and Irmeli Barkefors for their careful reading of
the manuscript. This work was supported by the European Research
Council (Grant no. 616047), Vetenskapsr{\aa}det, the Knut and Alice
Wallenberg Foundation, and the Swedish strategic research programme
eSSENCE.

%% file: contributions.tex
M.L., V.C. and J.E. designed the project. M.L. and V.C. designed and
implemented data analysis algorithms. E.A. designed and ran microscopy
experiments. V.C. and M.L. analyzed data. All authors wrote the paper.

%% file: osak_arxiv.bbl
%

%% file: supplementarytext.tex
\begin{figure}[b]
\centering
\includegraphics{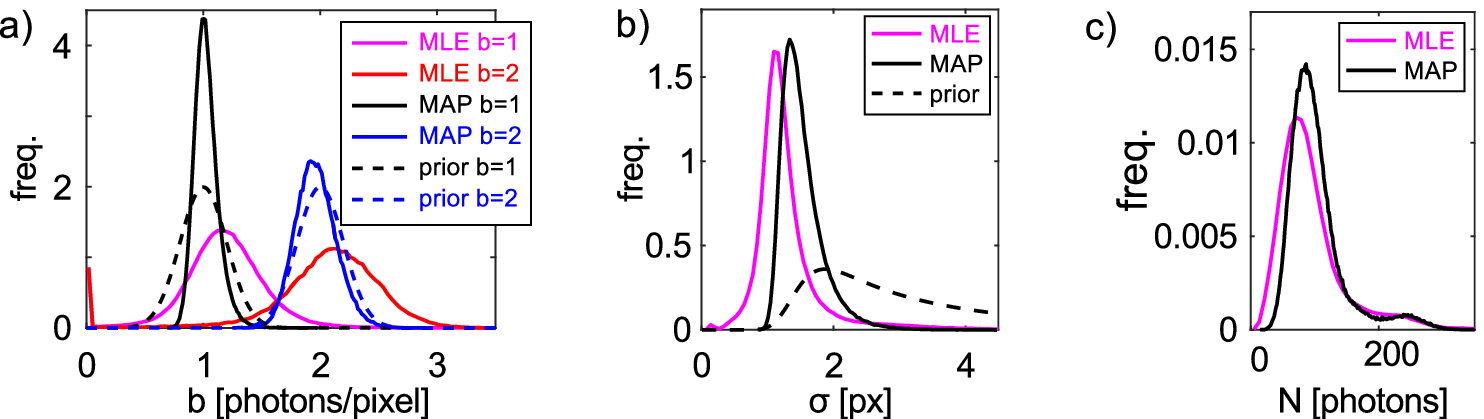}
\caption{\label{fig:parPrior} Prior distributions and fit parameter
  distributions corresponding to Fig.~\ref{MT:fig:simErr}a,c in the
  main text. (a) MLE and MAP background parameters, with $b=1$ and
  $b=2$ populations separated. Log-normal priors for the background
  are indicated by dashed lines. Note how the population near $b=0$ is
  eliminated in the MAP fits. (b) PSF width $\sigma$ for all
  backgrounds with (MAP) and without (MLE) priors, together with the
  skewed log-normal prior (dashed, reproduces
  Fig.~\ref{MT:fig:simErr}b). (c) Spot amplitude $N$
  distributions. Again, the population is shifted away from the very
  smallest amplitude values, even though no prior on $N$ was used in
  the MAP fit.}
\end{figure}
\section{Localization error estimates with different priors}\label{sec:locpriors}
In this section we present how different prior functions assigned to
the PSF width $\sigma$ and the background $b$ influence the accuracy
of the computed RMS errors and RMS uncertainties for the synthetic
data set considered in this paper.

We considered log-normal priors for the background intensity, either
centered around the true (simulated) background in each data set, or
around the average background intensity in all data sets.  For the PSF
width we tested two types of priors: (i) a log-normal prior around the
expected value for the PSF width; (ii) a skew-normal prior on
$\ln(\sigma)$, which assigns very low probability to PSF widths below
$\approx 0.8$ pixels, as well as low probabilities for very wide PSFs.
We do not impose any prior on the spot amplitude $N$.

The probability density function for the log-normal prior, denoted as
$\ln N(\mu,\sigma^2)$, is defined by
\begin{equation}
\label{eq:logNormal_prior}
f_{ln}(x,\sigma,\mu) = 
\frac{1}{x\sqrt{2\pi\sigma^2}}e^{-\frac{(ln x-\mu)^2}{2\sigma^2}},
\end{equation}
where $\mu$ and $\sigma$ are the mean and standard deviation
parameters. The skew-normal distribution \cite{azzalini1985} has a
density function given by
\begin{equation}
f_{sn}(x,x_0,\omega,\alpha) = 
\frac{2}{\omega}\phi\left(\frac{x-x_0}{\omega}\right)\Phi\left(\alpha\frac{x-x_0}{\omega}\right),
\end{equation}
where $\phi$ is the probability density function and $\Phi$ is the
cumulative distribution function for the unit normal distribution
$N(0,1)$, and the parameter $\alpha$ determines the degree of
asymmetry; with $\alpha$ positive (negative) increasing the weight
above (below) the mean value parameter $x_0$.

For our investigations on the effects of different prior
distributions, we used two background priors:
\begin{itemize}
\item
Log-normal ``correct'' prior on $b$: $\ln N(\ln b_\text{true},0.2^2)$,
where $b_\text{true}$ is either 1 or 2 photons/pixel, which is the
background level used for simulating that particular data set.
\item
Log-normal ``average'' prior on $b$: $\ln N(\ln 1.5,0.2^2)$, centered
around the average background intensity in all data sets, that is, 1.5
photons/pixel.
\end{itemize}
For the PSF width, we considered the two following priors:
\begin{itemize}
\item
Log-normal prior on $\sigma$: $\ln N(\ln 1.5,0.2^2)$, since the
half-width of the focused PSF width is around 1.5 pixels wide.
\item
Skew-normal prior on $\ln(\sigma)$: $f_{sn}(x,x_0,\omega,\alpha)$, where 
$x_0=\ln(1.5)$, $\alpha = 5$ and $\omega=1$.
\end{itemize}

\begin{figure*}
\centering
\includegraphics{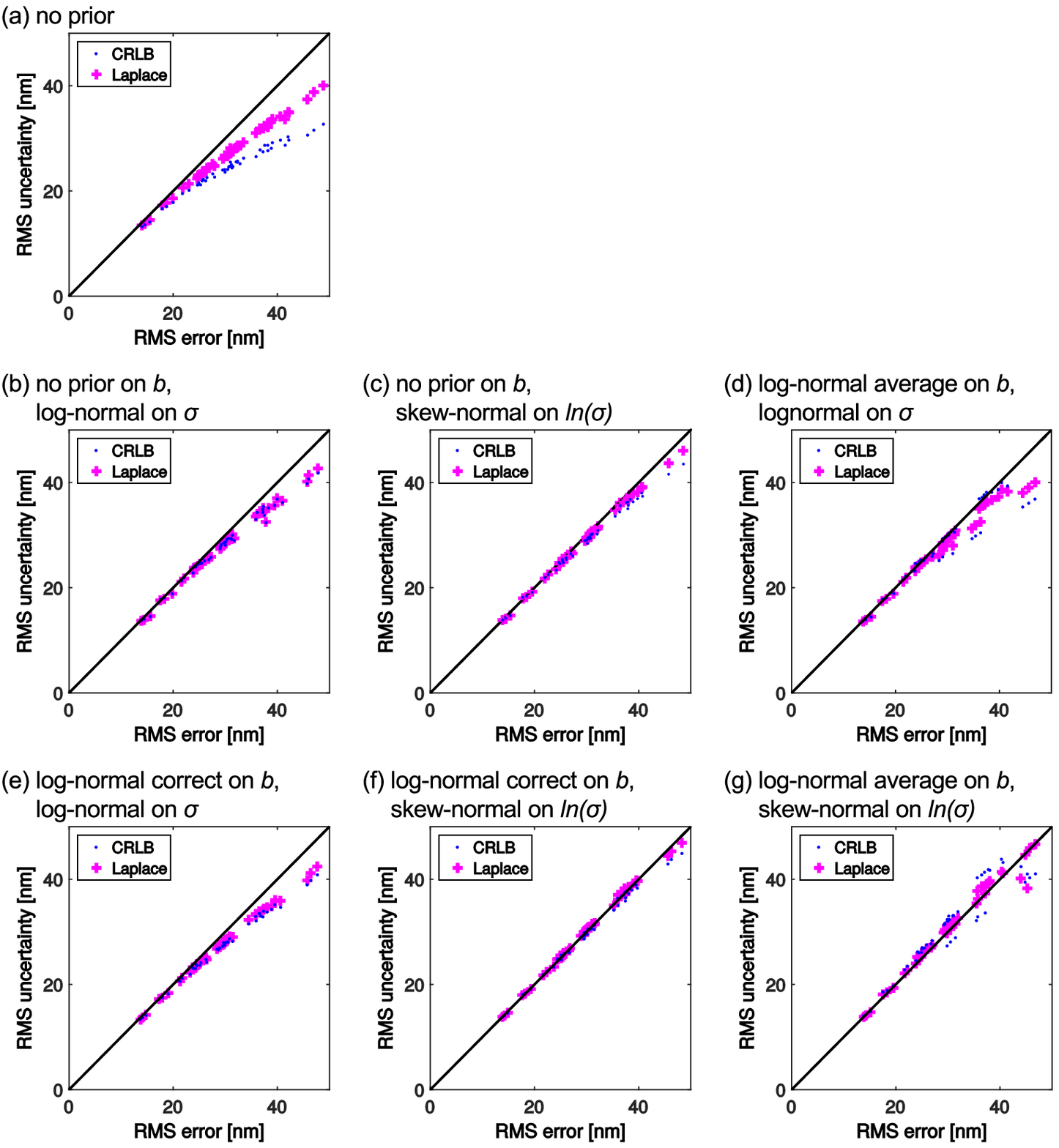}
\caption{\label{fig:priors_SI} Error-uncertainty plots for different
  combinations of prior distributions on $b$ and $\sigma$.}
\end{figure*}

Error-uncertainty comparisons for various combinations of the above
priors are shown in Fig.~\ref{fig:priors_SI}. As seen in
Fig.~\ref{fig:priors_SI}a (which reproduces Fig.~\ref{MT:fig:simErr}a
in the main text), imposing a flat prior (no prior) leads to estimated
uncertainties that substantially underestimate the true errors,
especially for points with low photon count.  In comparison, all other
prior combinations present an average improvement. Overall, the
Laplace estimator seems more consistent (RMS uncertainty mostly closer
to RMS error), and also less sensitive to the choice of priors. The
latter trend is especially clear in the examples using the ``average''
background prior, which is always either a systematic over- or
underestimate, which results in substantially larger variability or
the CRLB results.

\begin{figure}
\centering
\includegraphics{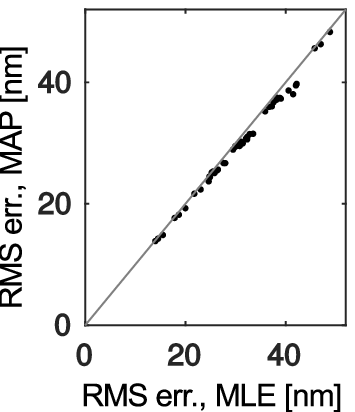}
\caption{\label{fig:MLE_vs_MAPerrors} Localization errors for MLE vs
  MAP localizations, using the prior described in the main text and
  Fig.~\ref{fig:priors_SI}f.}
\end{figure}

\section{Diffusion with time-varying localization errors}\label{sec:blurmodel}
Here, we give a detailed derivation of the various estimators and
models mentioned in the main text methods section, starting from
\Eq{MT:eq:blurdef}, the assumption that localizing a moving object
amounts to detecting the time-averaged position with some
(independent) localization error, which we will assume to be Gaussian.
Parts of these derivations have been given elsewhere
\cite{berglund2010,relich2016}, but are restated here in a different
form that facilitates a generalization to multi-state models.
\subsection{Diffusive camera-based tracking with blur and localization errors}
To streamline the presentation, we will use units where $\dt=1$, use
the subscript $t=0,1,2,\ldots$ to denote discrete time dependence, and
use the step length and localization variances, $\lambda_t=2D_t\dt$
and $v_t=\varepsilon_t^2$, respectively. We allow both of them to be
time-dependent, but assume them to be statistically independent and
restrict the diffusion constant to be constant throughout each
frame. Then, \Eq{MT:eq:blurdef} reads
\begin{equation}\label{eq:localization}
  x_t=\int_0^1f(t') y(t+t')dt'+\sqrt{v_t}\xi_t,
\end{equation}
where $\xi_t$ are independent identically distributed (iid) N(0,1)
variables, and $y(t)$ is the true trajectory of the particle being
localized. The shutter distribution $f(t)$ is a probability density on
$[0,1]$, which describes the image acquisition process (e.g., $f(t)=1$
for continuous acquisition), but neglects stochastic elements such as
fluorophore blinking. It has the distribution function
\begin{equation}
  F(t)=\int_0^t f(t')dt'.
\end{equation}

We divide $y(t)$ in two parts, the true positions $y_t$ at the
beginning of each frame, which evolve according to 
\begin{equation}\label{eq:ydiffusion}
  y_{t+1}=y_t+\sqrt{\lambda_t}\eta_t,
\end{equation}
where $\eta_t$ are again iid N(0,1), and a conditional interpolating
process between them, described by Brownian bridges
\cite{chow2009}. Thus, for $0\le t'\le 1$, we write
\begin{equation}\label{eq:interp}
  y(t+t')=y_t + t'(y_{t+1}-y_t)+\sqrt{\lambda_t}B_t(t'),
\end{equation}
where $B_t$ are a set of iid standard Brownian bridges. These are
Gaussian processes on the interval [0,1], defined by
\begin{equation}\label{eq:bridges}
  B_t(0)=B_t(1)=0,\quad 
  \mean{B_t(t')}=0,\quad
  \mean{B_t(t')B_t(t'')}=t'(1-t'')\text{, for $t'\le t''$},
\end{equation}
and also independent on different intervals, so that
$\mean{B_r(t')B_v(t'')}=0$ if $r\ne v$. Substituting the interpolation
formula \Eq{eq:interp} in the localization model \Eq{eq:localization},
we get
\begin{equation}\label{eq:interp2}
  x_t=y_t(1-\tau)+ y_{t+1}\tau+\sqrt{v_t}\xi_t
  +\sqrt{\lambda_t}\int_0^1f(t')B_t(t')dt',
\end{equation}
where we have introduced the shutter average, given by
\begin{equation}
  \tau=\int_0^1tf(t)dt.
\end{equation}
Using the properties of Brownian bridges, \Eq{eq:bridges}, one can
show that the last integral in \Eq{eq:interp2} is a Gaussian random
variable with mean zero and variance
\begin{equation}
  \beta\equiv\Var\left[\int_0^1f(t')B_t(t')dt'\right]=\tau(1-\tau)-R,
\end{equation}
where $R$ is the blur coefficient of Ref.~\cite{berglund2010}, given
by
\begin{equation}
  R=\int_0^1F(t)\big(1-F(t)\big)dt.
\end{equation}
By assumption, $v_t$ and $B_t$ are statistically independent, and thus
one can add up the noise in the measurement model and arrive at 
\begin{equation}\label{eq:interp3}
  x_t=y_t(1-\tau)+ y_{t+1}\tau+\sqrt{v_t+\beta\lambda_t}\zeta_t,
\end{equation}
where $\zeta_t$ are again iid N(0,1). 

\subsection{Constant exposure}
An important class of shutter distribution are those that are constant
during some fraction $t_E$ of the each frame, and then zero, i.e.,
\begin{equation}
  f(t)=\left\{\begin{array}{ll}
  \frac{1}{t_E},&t\le t_E,\\
  0,&t> t_E,\\
  \end{array}\right.
\end{equation}
which leads to
\begin{equation}
  \tau=\frac{t_E}{2},\quad
  R=\frac{t_E}{6},\quad
  \beta=\frac14 t_E(\frac43-t_E).
\end{equation}
We see that $R\le \frac16$ and $\tau\le\frac12$, with maxima at
continuous exposure ($t_E=1$).  

On the other hand, $\beta$ has a maximum of $\frac19$ at
$t_E=\frac23$, and the value of $\frac{1}{12}$ at continuous exposure
can only be further lowered when $t_E<\frac13$.

It is unclear if this is significant, since with non-constant exposure
there is some freedom in how the shutter distribution is defined, and
one could also place it symmetrically in the interval and get
$\tau=0.5$ for all exposure times. We defer further investigations of
this issue to future work.

\subsection{Covariance relations}\label{sec:cve}
The covariance matrix for the steps $\Delta_t=x_{t+1}-x_t$ can be
found from Eqs.~(\ref{eq:ydiffusion},\ref{eq:interp3}). With some
manipulations, we get
\begin{align}
  \mean{\Delta_t^2}= &(1-\tau)\lambda_t+\tau\lambda_{t+1}
  -(\lambda_{t+1}+\lambda_t)R+v_t+v_{t+1},\\
  \mean{\Delta_t\Delta_{t+1}}= &\lambda_{t+1}R-v_{t+1},\\
  \mean{\Delta_t\Delta_{t+t'}}= &0\text{, if $|t'|>1$},
\end{align}
where the expectations $\mean{\cdot}$ are understood to be over the
noise distributions only. If we further assume simple diffusion,
$\lambda_t=2D\dt=const.$, and average over time as well, we recover
covariance relations of the same form as \Eq{MT:eq:dxcov},
\begin{equation}
  \mean{\Delta_t^2}=2D\dt(1-2R)+2\mean{v_t},\quad
  \mean{\Delta_t\Delta_{t\pm 1}}=2D\dt R-\mean{v_t},
\end{equation}
where the averages are now over time as well, and $\varepsilon^2$ is
identified as the time-average $\mean{v_t}$. Thus, the
covariance-based estimators of Ref.~\cite{vestergaard2014} should apply
also to non-constant localization errors.
\section{Maximum likelihood estimator}\label{sec:diffusionMLE}
For the case of a single trajectory of simple diffusion in one
dimension, the likelihood for the diffusion constant follows from
Eqs.~\ref{eq:ydiffusion} and \ref{eq:interp3},
\begin{eqnarray}
  p(x|\lambda)  &=& \int dy p(x|y,\lambda)p(y|\lambda),\label{eq:lamlik}\\
  p(y|\lambda)  &=&\prod_{t=1}^T \big(2\pi\lambda\big)^{-\frac12}
  \exp\Big[-\frac{(y_{t+1}-y_t)^2}{2\lambda}\Big],\\
  p(x|y,\lambda)&=& \prod_{t=1}^T \big(2\pi(v_t+\beta\lambda)\big)^{-\frac12}
  \exp\Big[-\frac{1}{2}(v_t+\beta\lambda)^{-1}
    \big(x_t-(1-\tau)y_t-\tau y_{t+1}\big)^2\Big],
\end{eqnarray}
where we neglected to supply a starting density for $y_1$, since the
problem is translation invariant. The integral over the hidden path in
\Eq{eq:lamlik} is a multivariate Gaussian and can be solved exactly in
several ways \cite{relich2016}. Defining
\begin{equation}
  \vec{y}=[y_1,\;y_2,\ldots,y_{T+1}]^\dagger,\quad
  \vec{x}=[x_1,\;x_2,\ldots,x_{T}]^\dagger,
\end{equation}
where $^\dagger$ denotes matrix transpose, we get
\begin{multline}\label{eq:pxlam1}
  p(x|\lambda)=\int d\vec{y} \exp\bigg[
   -T\ln(2\pi)-\frac{T}{2}\ln(\lambda)
   -\frac12\sum_{t=1}^T\ln(v_t+\beta\lambda)
   -\frac12\Big(\vec{y}^\dagger\Lambda\vec{y}-2\vec{y}^\dagger W\vec{x}
   +\vec{x}^\dagger V\vec{x}
    \Big)\bigg]\\
  =\exp\bigg[
   -T\ln(2\pi)-\frac{T}{2}\ln(\lambda)
   -\frac12\sum_{t=1}^T\ln(v_t+\beta\lambda)\bigg]\\
  \times\int d\vec{y}\exp\bigg[
    -\frac12\Big(
    (\vec{y}-\Lambda^{-1}W\vec{x})^\dagger\Lambda(\vec{y}-\Lambda^{-1}W\vec{x})    
   +\vec{x}^\dagger (V-W^\dagger\Lambda^{-1} W)\vec{x}
    \Big)\bigg],
\end{multline}
where $\Lambda,W,V$ are matrices whose \vc{elements} are found by
comparing terms.  This is a multivariate Gaussian in $\vec{y}$, with
mean value $\mu=\Lambda^{-1}W\vec{x}$ and covariance matrix
$\Sigma=\Lambda^{-1}$, and the marginalized likelihood is therefore
given by
\begin{equation}\label{eq:pxlam2}
  p(x|\lambda)=
\exp\bigg[
   -\frac{T-1}{2}\ln(2\pi)-\frac{T}{2}\ln(\lambda)
   -\frac12\sum_{t=1}^T\ln(v_t+\beta\lambda) 
   -\frac12\vec{x}^\dagger (V-W^\dagger\Lambda^{-1}W)\vec{x} 
   -\frac12\ln|\Lambda|\bigg].
\end{equation}
By comparing terms, we see that $\Lambda$ is symmetric tridiagonal and
positive definite, $V$ is diagonal, and $W$ has only non-zero elements
on the diagonal and first upper diagonal, and so it is possible to
compute the above matrix expression in linear time. We minimize $\ln
p(x|\lambda)$ by standard optimization routines in Matlab.

\begin{figure}
\centering
  \includegraphics{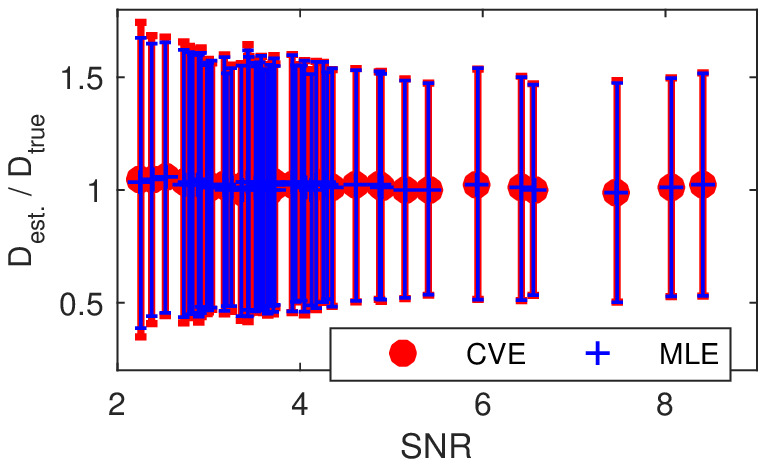}
  \caption{\label{fig:DestMLE} Mean value and 1\% quantiles of estimated
diffusion constants using the CVE and MLE diffusion constant
estimator, both using estimated positions and uncertainties.}
  
\end{figure}

Repeating the analysis of Fig.~\ref{MT:fig:Dest} in the main text for
this MLE estimator, see Fig.~\ref{fig:DestMLE}, we find almost no
improvement over the CVE that uses average estimated
uncertainties. However, the fact that the MLE estimator is limited to
return positive estimated diffusion constants, and handles missing
data points (by assigning them infinite or very large uncertainties),
may be useful in applications.

\section{Variational EM algorithm for diffusive HMM}\label{sec:HMM}
Here, we extend the above diffusion model to include multiple
diffusion constants. We start by writing down a diffusive HMM, which
includes both the hidden path of the above diffusion model, but also a
set of hidden states with different diffusion constants, that evolve
as a discrete Markov process. Similar models (that however did not
include explicit motion blur effects), have previously been solved by
stochastic EM algorithms \cite{ashley2015,bernstein2016}. Here, we
instead describe a deterministic variational approach inspired by
algorithms for factorial HMMs \cite{ghahramani1997}.

In the rest of this section, we proceed as follows: We start by
specify the diffusive HMM model for a single 1-dimensional
trajectory. We then outline the variational EM approach, derive high
level update equations, and describe the procedure for re-estimating
localized positions. However, do not give a detailed derivation of all
steps in the algorithm, as large parts of it closely resembles
previously published derivations of variational algorithms for HMMs
\cite{persson2013,johnson2014}.

\subsection{Model}
In addition to the measured ($\vec{x}$) and true ($\vec{y}$)
positions, we include a hidden state trajectory
$\vec{s}=[s_1,s_2,\ldots,s_T]$, such that $s_t$ determines the
diffusion constant on the interval $[t,t+1]$. The hidden states are
numbered from 1 to $N$, and evolve according to a Markov process with
transition matrix $A$ and initial state probability $\vpp$.  For a
single 1-dimensional trajectory, this leads to a complete data
likelihood of the form
\begin{equation}
  p(\x,\y,\s|\lambda,A,\vpp)= p(\x|\y,\s,\lambda)p(\y|\s,\lambda)p(\s|A,\vpp),
\end{equation}
with factors
\begin{eqnarray}
  p(\s|A,\vpp)&=&
  \prod_{m=1}^N\pp_m^{\delta_{m,s_1}}
  \prod_{t=2}^T\prod_{i,j=1}^N A_{ij}^{\delta_{is_t}\delta_{js_{t+1}}},\\
  p(\y|\s,\lambda)&=&
  \prod_{t=1}^T\prod_{j=1}^N (2\pi \lambda_j)^{-\frac12\delta_{js_t}}
  \exp\left[-\delta_{js_t}\frac{(y_{t+1}-y_t)^2}{2\lambda_j}\right],\\
  p(\x|\y,\s,\lambda)&=&
  \prod_{t=1}^T\prod_{j=1}^N \big(2\pi(v_t+\beta\lambda_j)\big)^{-\frac12\delta_{js_t}}
  \exp\left[-\delta_{js_t}
    \frac{(x_t-(t-\tau)y_t-\tau y_{t+1})^2}{2(v_t+\beta\lambda_j)}
    \right].
\end{eqnarray}

\subsection{Variational EM approach} 
We would like to perform maximum-likelihood inference of the model
parameters, which means maximizing the likelihood with latent
variables $\s,\y$ integrated out,
\begin{equation}
  L(A,\vpp,\lambda)=\int d\y\sum_{\s}
  p(\x|\y,\s,\lambda)p(\y|\s,\lambda)p(\s|A,\vpp).
\end{equation}
Since this problem is intractable, we make a variational
approximation, meaning we approximate $\ln L$ with a lower bound
\begin{equation}\label{eq:lowerbound}
  \ln L =\ln \int d\y\sum_{\s}
  p(\x|\y,\s,\lambda)p(\y|\s,\lambda)p(\s|A,\vpp)
  \ge \int d\y\sum_{\s} q(\s)q(\y)\ln
  \frac{p(\x|\y,\s,\lambda)p(\y|\s,\lambda)p(\s|A,\vpp)}{q(\s)q(\y)}\equiv F,
\end{equation}
where the inequality follows from Jensen's inequality.  Here, $q(\s)$,
$q(\y)$ are arbitrary variational distributions that need to be
optimized together with the model parameters to achieve the tightest
lower bound, and can be used for approximate inference about the
latent variables. In particular, it turns out that the optimal
variational distributions approximate the posterior distribution of
$\y,\s$ in the sense of minimizing a Kullback-Leibler divergence
\cite{bishop2006}.  We use gradient descent for the optimization,
i.e., we iteratively optimize each variational distribution and the
parameters with the others fixed, which leads to an EM-type algorithm.

To optimize $F$ w.r.t.~the variational distributions, we set the
functional derivatives of $F$ to zero and use Lagrange multipliers to
enforce normalization. After some work, one arrives at
\begin{eqnarray}
  \ln q(\s)&=&-\ln Z_s
  +\mean{\ln p(\x|\y,\s,\lambda}_{q(\y)}
  +\mean{\ln p(\y|\s,\lambda}_{q(\y)}
  +\ln p(\s|A,\vpp),\label{eq:lnqs}\\
  \ln q(\y)&=&-\ln Z_y
  +\mean{\ln p(\x|\y,\s,\lambda}_{q(\s)}
  +\mean{\ln p(\y|\s,\lambda}_{q(\s)}
  +\underbrace{\mean{\ln p(\s|A,\vpp)}_{q(\s)}}_\text{independent of $\y$},
\end{eqnarray}
where $Z_{s,y}$ are normalization constants originating from the
Lagrange multipliers, and $\mean{\cdot}_{f}$ denotes an expectation
value computed with respect to the distribution $f$. As it turns out,
these equations are individually tractable, and results in $q(\y)$
being a multivariate Gaussian, and $q(\s)$ adopting the standard HMM
form amenable to efficient forward-backward iterations.  For the
initial state and transition probability parameters, the only
dependence in $F$ comes from $p(\s|A,\lambda)$, and arrives at the
classical Baum-Welch reestimation formulas \cite{rabiner1986}.
However, optimizing $F$ w.r.t.~step length variances does not lead to
a tractable update equation, and we are instead forced to optimize the
$\lambda_j$-dependent parts of $F$ numerically, i.e.,
\begin{equation}
  \lambda_j=\mathrm{argmax}_{\lambda_j}
  \mean{\ln p(\x|\y\s,\lambda)}_{q(\y)q(\s)}
  \mean{\ln p(\y|\s,\lambda)}_{q(\y)q(\s)},
\end{equation}
\subsection{The lower bound}
The lower bound can be computed as well, and gets a
particularly simple form just after the update of
$q(\s)$. Substituting the update equation \Eq{eq:lnqs} into the
expression for $F$, \Eq{eq:lowerbound}, we get
\begin{multline}
  F=\int d\y\sum_{\s} q(\s)q(\y)\ln\Big[
\ln p(\x|\y,\s,\lambda)+\ln p(\y|\s,\lambda)+\ln p(\s,A,\vpp)\\
+\ln Z_s
-\mean{\ln p(\x|\y,\s,\lambda)+\ln p(\y|\s,\lambda)+\ln p(\s,A,\vpp)}_{q(\y)}
-\ln q(\y)\Big]
  =\ln Z_s-\mean{\ln q(\y)}_{q(\y)}.
\end{multline}
Here, $\ln Z_s$ is the normalization constant of $q(\s)$ that can be
computed as part of the forward-backward iteration, and since (as we
noted above) $q(\y)$ is a multivariate Gaussian with dimension $T+1$,
we get
\begin{equation}
  -\mean{\ln q(\y)}_{q(\y)}=
  \frac{d}{2}(T+1)\big(1+\ln(2\pi)\big)+\frac12\ln|\Sigma|,
\end{equation}
where $\Sigma$ is the covariance matrix of $q(\y)$.  The inverse of
$\Sigma$ is analogous to the matrix $\Lambda$ appearing in the
single-state diffusion estimator,
Eqs.~(\ref{eq:pxlam1},\ref{eq:pxlam2}), and in particular
$\Sigma^{-1}$ is also symmetric, tridiagonal, and positive definite,
and thus the determinant $\ln|\Sigma|=-\ln|\Sigma^{-1}|$ can be
robustly computed in linear time. Since the EM algorithm approximates
the parameter likelihood, the lower bound cannot be used for model
selection as in the case of variational maximum evidence calculations
\cite{bronson2009,persson2013,johnson2014}. However, it is still
useful for numerical convergence control, and possibly for model
selection together with some complexity penalty such as the Bayesian
or Akaike information criterion \cite{burnham}.

\subsection{Refined positions}\label{sec:positions}
An additional use for the HMM analysis is to use it to refine the
localizations. Since the HMM pools information about many spots, it is
in principle possible to beat the Cram\'er-Rao lower bound for single
image localizations in this way.  To set up the refinment problem, we
refer back to \Eq{SI:eq:interp2} and the different contributions to
the observed position $x_t=z_t+\sqrt{v_t}\xi_t$, where
\begin{equation}\label{eq:zrefine1}
  z_t=y_t(1-\tau)+ y_{t+1}\tau+\sqrt{\lambda_{s_t}}\int_0^1f(t')B_t(t')dt'
\end{equation}
is the motion-averaged position that the localization algorithm tries
to estimate (according to this model). To refine the localization, we
need to compute the posterior density of $z_t$, i.e.,
$p(z_t|\x,\theta)$.

We start by recalling that the Brownian bridge integral in
\Eq{SI:eq:zrefine1} is Gaussian with mean zero and variance
$\beta\lambda_{s_t}$, and using the compact notation
$\overline{y_t}=(1-\tau)y_t+\tau y_{t+1}$ and $\theta$ to denote all
the model parameters, we therefore have
\begin{equation}
  p(z_t|\y,\s,\theta)=p(z_t|\overline y_t,s_t,\theta)
  =N(\overline y_t,\beta\lambda_{s_t}),
\end{equation}
where $N(a,b)$ denotes a Gaussian density with mean $a$ and variance
$b$.  Furthermore, the localization uncertainty is also assumed
Gaussian and independent of the underlying kinetics:
\begin{equation}
  p(x_t|\z,\y,\s,\theta)=p(x_t|z_t)=N(z_t,v_t).
\end{equation}
Applying Bayes theorem to these relations, we get
\begin{equation}
  p(z_t|x_t,\y,\s,\theta)
  =
  \frac{p(x_t|z_t)p(z_t|\y,\s,\theta)}{p(x_t|\y,\s,\theta))}
  =\frac{N(z_t,v_t)N(\overline y_t,\beta\lambda_{s_t})}{
    N(\overline y_t,\beta\lambda_{s_t}+v_t)}
  =
  N\Big(\frac{v_t\overline y_t+\beta\lambda_{s_t}x_t}{v_t+\beta\lambda_{s_t}},
  \frac{\beta\lambda_{s_t}v_t}{v_t+\beta\lambda_{s_t}}
  \Big).
\end{equation}
The predictive distribution is finally given by marginalizing
$p(z_t|x_t,\y,\s,\theta)$ over the posterior for $\y,\s$. Using the
variational distribution, this means
\begin{equation}
  p(z_t|\x,\theta)
  \approx\mean{
  N\Big(\frac{v_t\overline y_t+\beta\lambda_{s_t}x_t}{v_t+\beta\lambda_{s_t}},
  \frac{\beta\lambda_{s_t}v_t}{v_t+\beta\lambda_{s_t}}\Big)}_{q(\y)q(\s)}.
\end{equation}
In particular, the posterior mean of $z_t$ is then given by
\begin{equation}
  \mean{z_t|\x,\theta}\approx\mean{
    \frac{v_t\overline y_t+\beta\lambda_{s_t}x_t}{v_t+\beta\lambda_{s_t}}
  }_{q(\y)q(\s)}
=\mean{
    \frac{(1-\tau)\mu_t+\tau\mu_{t+1}
      +\frac{\beta\lambda_{s_t}}{v_t}x_t}{1+\frac{\beta\lambda_{s_t}}{v_t}}
  }_{q(\s)},
\end{equation}
which we will use as our estimator for refining the localizations.
Here, $\mu_t=\mean{y_t}_{q(\y)}$ is the variational mean value.  The
variational average over hidden states is done numerically.

%% file: osak_arxiv.bbl
\begin{thebibliography}{37}%
\makeatletter
\providecommand \@ifxundefined [1]{%
 \@ifx{#1\undefined}
}%
\providecommand \@ifnum [1]{%
 \ifnum #1\expandafter \@firstoftwo
 \else \expandafter \@secondoftwo
 \fi
}%
\providecommand \@ifx [1]{%
 \ifx #1\expandafter \@firstoftwo
 \else \expandafter \@secondoftwo
 \fi
}%
\providecommand \natexlab [1]{#1}%
\providecommand \enquote  [1]{``#1''}%
\providecommand \bibnamefont  [1]{#1}%
\providecommand \bibfnamefont [1]{#1}%
\providecommand \citenamefont [1]{#1}%
\providecommand \href@noop [0]{\@secondoftwo}%
\providecommand \href [0]{\begingroup \@sanitize@url \@href}%
\providecommand \@href[1]{\@@startlink{#1}\@@href}%
\providecommand \@@href[1]{\endgroup#1\@@endlink}%
\providecommand \@sanitize@url [0]{\catcode `\\12\catcode `\$12\catcode
  `\&12\catcode `\#12\catcode `\^12\catcode `\_12\catcode `\%12\relax}%
\providecommand \@@startlink[1]{}%
\providecommand \@@endlink[0]{}%
\providecommand \url  [0]{\begingroup\@sanitize@url \@url }%
\providecommand \@url [1]{\endgroup\@href {#1}{\urlprefix }}%
\providecommand \urlprefix  [0]{URL }%
\providecommand \Eprint [0]{\href }%
\providecommand \doibase [0]{http://dx.doi.org/}%
\providecommand \selectlanguage [0]{\@gobble}%
\providecommand \bibinfo  [0]{\@secondoftwo}%
\providecommand \bibfield  [0]{\@secondoftwo}%
\providecommand \translation [1]{[#1]}%
\providecommand \BibitemOpen [0]{}%
\providecommand \bibitemStop [0]{}%
\providecommand \bibitemNoStop [0]{.\EOS\space}%
\providecommand \EOS [0]{\spacefactor3000\relax}%
\providecommand \BibitemShut  [1]{\csname bibitem#1\endcsname}%
\let\auto@bib@innerbib\@empty
\bibitem [{\citenamefont {Sage}\ \emph {et~al.}(2015)\citenamefont {Sage},
  \citenamefont {Kirshner}, \citenamefont {Pengo}, \citenamefont {Stuurman},
  \citenamefont {Min}, \citenamefont {Manley},\ and\ \citenamefont
  {Unser}}]{sage2015}%
  \BibitemOpen
  \bibfield  {author} {\bibinfo {author} {\bibfnamefont {D.}~\bibnamefont
  {Sage}}, \bibinfo {author} {\bibfnamefont {H.}~\bibnamefont {Kirshner}},
  \bibinfo {author} {\bibfnamefont {T.}~\bibnamefont {Pengo}}, \bibinfo
  {author} {\bibfnamefont {N.}~\bibnamefont {Stuurman}}, \bibinfo {author}
  {\bibfnamefont {J.}~\bibnamefont {Min}}, \bibinfo {author} {\bibfnamefont
  {S.}~\bibnamefont {Manley}}, \ and\ \bibinfo {author} {\bibfnamefont
  {M.}~\bibnamefont {Unser}},\ }\bibfield  {title} {\enquote {\bibinfo {title}
  {Quantitative evaluation of software packages for single-molecule
  localization microscopy},}\ }\href {\doibase 10.1038/nmeth.3442} {\bibfield
  {journal} {\bibinfo  {journal} {Nat. Meth.}\ }\textbf {\bibinfo {volume} {5}}
  (\bibinfo {year} {2015}),\ 10.1038/nmeth.3442}\BibitemShut {NoStop}%
\bibitem [{\citenamefont {Thompson}\ \emph {et~al.}(2002)\citenamefont
  {Thompson}, \citenamefont {Larson},\ and\ \citenamefont
  {Webb}}]{thompson2002}%
  \BibitemOpen
  \bibfield  {author} {\bibinfo {author} {\bibfnamefont {Russell~E}\
  \bibnamefont {Thompson}}, \bibinfo {author} {\bibfnamefont {Daniel~R}\
  \bibnamefont {Larson}}, \ and\ \bibinfo {author} {\bibfnamefont {Watt~W}\
  \bibnamefont {Webb}},\ }\bibfield  {title} {\enquote {\bibinfo {title}
  {Precise nanometer localization analysis for individual fluorescent
  probes},}\ }\href@noop {} {\bibfield  {journal} {\bibinfo  {journal}
  {Biophys. J.}\ }\textbf {\bibinfo {volume} {82}},\ \bibinfo {pages}
  {2775--2783} (\bibinfo {year} {2002})}\BibitemShut {NoStop}%
\bibitem [{\citenamefont {Ober}\ \emph {et~al.}(2004)\citenamefont {Ober},
  \citenamefont {Ram},\ and\ \citenamefont {Ward}}]{ober2004}%
  \BibitemOpen
  \bibfield  {author} {\bibinfo {author} {\bibfnamefont {Raimund~J}\
  \bibnamefont {Ober}}, \bibinfo {author} {\bibfnamefont {Sripad}\ \bibnamefont
  {Ram}}, \ and\ \bibinfo {author} {\bibfnamefont {E~Sally}\ \bibnamefont
  {Ward}},\ }\bibfield  {title} {\enquote {\bibinfo {title} {Localization
  accuracy in single-molecule microscopy},}\ }\href@noop {} {\bibfield
  {journal} {\bibinfo  {journal} {Biophys. J.}\ }\textbf {\bibinfo {volume}
  {86}},\ \bibinfo {pages} {1185--1200} (\bibinfo {year} {2004})}\BibitemShut
  {NoStop}%
\bibitem [{\citenamefont {Mortensen}\ \emph {et~al.}(2010)\citenamefont
  {Mortensen}, \citenamefont {Churchman}, \citenamefont {Spudich},\ and\
  \citenamefont {Flyvbjerg}}]{mortensen2010}%
  \BibitemOpen
  \bibfield  {author} {\bibinfo {author} {\bibfnamefont {Kim~I.}\ \bibnamefont
  {Mortensen}}, \bibinfo {author} {\bibfnamefont {L.~Stirling}\ \bibnamefont
  {Churchman}}, \bibinfo {author} {\bibfnamefont {James~A.}\ \bibnamefont
  {Spudich}}, \ and\ \bibinfo {author} {\bibfnamefont {Henrik}\ \bibnamefont
  {Flyvbjerg}},\ }\bibfield  {title} {\enquote {\bibinfo {title} {Optimized
  localization analysis for single-molecule tracking and super-resolution
  microscopy},}\ }\href {\doibase 10.1038/nmeth.1447} {\bibfield  {journal}
  {\bibinfo  {journal} {Nat. Meth.}\ }\textbf {\bibinfo {volume} {7}},\
  \bibinfo {pages} {377--381} (\bibinfo {year} {2010})}\BibitemShut {NoStop}%
\bibitem [{\citenamefont {Rieger}\ and\ \citenamefont
  {Stallinga}(2014)}]{rieger2014}%
  \BibitemOpen
  \bibfield  {author} {\bibinfo {author} {\bibfnamefont {Bernd}\ \bibnamefont
  {Rieger}}\ and\ \bibinfo {author} {\bibfnamefont {Sjoerd}\ \bibnamefont
  {Stallinga}},\ }\bibfield  {title} {\enquote {\bibinfo {title} {The lateral
  and axial localization uncertainty in super-resolution light microscopy},}\
  }\href {\doibase 10.1002/cphc.201300711} {\bibfield  {journal} {\bibinfo
  {journal} {ChemPhysChem}\ }\textbf {\bibinfo {volume} {15}},\ \bibinfo
  {pages} {664--670} (\bibinfo {year} {2014})}\BibitemShut {NoStop}%
\bibitem [{\citenamefont {Betzig}\ \emph {et~al.}(2006)\citenamefont {Betzig},
  \citenamefont {Patterson}, \citenamefont {Sougrat}, \citenamefont
  {Lindwasser}, \citenamefont {Olenych}, \citenamefont {Bonifacino},
  \citenamefont {Davidson}, \citenamefont {Lippincott-Schwartz},\ and\
  \citenamefont {Hess}}]{betzig2006}%
  \BibitemOpen
  \bibfield  {author} {\bibinfo {author} {\bibfnamefont {Eric}\ \bibnamefont
  {Betzig}}, \bibinfo {author} {\bibfnamefont {George~H}\ \bibnamefont
  {Patterson}}, \bibinfo {author} {\bibfnamefont {Rachid}\ \bibnamefont
  {Sougrat}}, \bibinfo {author} {\bibfnamefont {O~Wolf}\ \bibnamefont
  {Lindwasser}}, \bibinfo {author} {\bibfnamefont {Scott}\ \bibnamefont
  {Olenych}}, \bibinfo {author} {\bibfnamefont {Juan~S}\ \bibnamefont
  {Bonifacino}}, \bibinfo {author} {\bibfnamefont {Michael~W}\ \bibnamefont
  {Davidson}}, \bibinfo {author} {\bibfnamefont {Jennifer}\ \bibnamefont
  {Lippincott-Schwartz}}, \ and\ \bibinfo {author} {\bibfnamefont {Harald~F}\
  \bibnamefont {Hess}},\ }\bibfield  {title} {\enquote {\bibinfo {title}
  {Imaging intracellular fluorescent proteins at nanometer resolution},}\
  }\href@noop {} {\bibfield  {journal} {\bibinfo  {journal} {Science}\ }\textbf
  {\bibinfo {volume} {313}},\ \bibinfo {pages} {1642--1645} (\bibinfo {year}
  {2006})}\BibitemShut {NoStop}%
\bibitem [{\citenamefont {Rust}\ \emph {et~al.}(2006)\citenamefont {Rust},
  \citenamefont {Bates},\ and\ \citenamefont {Zhuang}}]{rust2006}%
  \BibitemOpen
  \bibfield  {author} {\bibinfo {author} {\bibfnamefont {Michael~J}\
  \bibnamefont {Rust}}, \bibinfo {author} {\bibfnamefont {Mark}\ \bibnamefont
  {Bates}}, \ and\ \bibinfo {author} {\bibfnamefont {Xiaowei}\ \bibnamefont
  {Zhuang}},\ }\bibfield  {title} {\enquote {\bibinfo {title}
  {Sub-diffraction-limit imaging by stochastic optical reconstruction
  microscopy {(STORM)}},}\ }\href@noop {} {\bibfield  {journal} {\bibinfo
  {journal} {Nat. Meth.}\ }\textbf {\bibinfo {volume} {3}},\ \bibinfo {pages}
  {793--796} (\bibinfo {year} {2006})}\BibitemShut {NoStop}%
\bibitem [{\citenamefont {Lind\'{e}n}\ \emph {et~al.}(2016)\citenamefont
  {Lind\'{e}n}, \citenamefont {\'{C}uri\'{c}}, \citenamefont {Boucharin},
  \citenamefont {Fange},\ and\ \citenamefont {Elf}}]{linden2016}%
  \BibitemOpen
  \bibfield  {author} {\bibinfo {author} {\bibfnamefont {Martin}\ \bibnamefont
  {Lind\'{e}n}}, \bibinfo {author} {\bibfnamefont {Vladimir}\ \bibnamefont
  {\'{C}uri\'{c}}}, \bibinfo {author} {\bibfnamefont {Alexis}\ \bibnamefont
  {Boucharin}}, \bibinfo {author} {\bibfnamefont {David}\ \bibnamefont
  {Fange}}, \ and\ \bibinfo {author} {\bibfnamefont {Johan}\ \bibnamefont
  {Elf}},\ }\bibfield  {title} {\enquote {\bibinfo {title} {Simulated single
  molecule microscopy with {SMeagol}},}\ }\href {\doibase
  10.1093/bioinformatics/btw109} {\bibfield  {journal} {\bibinfo  {journal}
  {Bioinformatics}\ } (\bibinfo {year} {2016}),\
  10.1093/bioinformatics/btw109},\ \bibinfo {note}
  {\url{http://smeagol.sourceforge.net}}\BibitemShut {NoStop}%
\bibitem [{\citenamefont {Manley}\ \emph {et~al.}(2008)\citenamefont {Manley},
  \citenamefont {Gillette}, \citenamefont {Patterson}, \citenamefont {Shroff},
  \citenamefont {Hess}, \citenamefont {Betzig},\ and\ \citenamefont
  {Lippincott-Schwartz}}]{manley2008}%
  \BibitemOpen
  \bibfield  {author} {\bibinfo {author} {\bibfnamefont {Suliana}\ \bibnamefont
  {Manley}}, \bibinfo {author} {\bibfnamefont {Jennifer~M.}\ \bibnamefont
  {Gillette}}, \bibinfo {author} {\bibfnamefont {George~H.}\ \bibnamefont
  {Patterson}}, \bibinfo {author} {\bibfnamefont {Hari}\ \bibnamefont
  {Shroff}}, \bibinfo {author} {\bibfnamefont {Harald~F.}\ \bibnamefont
  {Hess}}, \bibinfo {author} {\bibfnamefont {Eric}\ \bibnamefont {Betzig}}, \
  and\ \bibinfo {author} {\bibfnamefont {Jennifer}\ \bibnamefont
  {Lippincott-Schwartz}},\ }\bibfield  {title} {\enquote {\bibinfo {title}
  {High-density mapping of single-molecule trajectories with photoactivated
  localization microscopy},}\ }\href {\doibase 10.1038/nmeth.1176} {\bibfield
  {journal} {\bibinfo  {journal} {Nat. Meth.}\ }\textbf {\bibinfo {volume}
  {5}},\ \bibinfo {pages} {155--157} (\bibinfo {year} {2008})}\BibitemShut
  {NoStop}%
\bibitem [{\citenamefont {Vestergaard}\ \emph {et~al.}(2014)\citenamefont
  {Vestergaard}, \citenamefont {Blainey},\ and\ \citenamefont
  {Flyvbjerg}}]{vestergaard2014}%
  \BibitemOpen
  \bibfield  {author} {\bibinfo {author} {\bibfnamefont {Christian~L.}\
  \bibnamefont {Vestergaard}}, \bibinfo {author} {\bibfnamefont {Paul~C.}\
  \bibnamefont {Blainey}}, \ and\ \bibinfo {author} {\bibfnamefont {Henrik}\
  \bibnamefont {Flyvbjerg}},\ }\bibfield  {title} {\enquote {\bibinfo {title}
  {Optimal estimation of diffusion coefficients from single-particle
  trajectories},}\ }\href {\doibase 10.1103/PhysRevE.89.022726} {\bibfield
  {journal} {\bibinfo  {journal} {Phys. Rev. E}\ }\textbf {\bibinfo {volume}
  {89}},\ \bibinfo {pages} {022726} (\bibinfo {year} {2014})}\BibitemShut
  {NoStop}%
\bibitem [{\citenamefont {Relich}\ \emph {et~al.}(2016)\citenamefont {Relich},
  \citenamefont {Olah}, \citenamefont {Cutler},\ and\ \citenamefont
  {Lidke}}]{relich2016}%
  \BibitemOpen
  \bibfield  {author} {\bibinfo {author} {\bibfnamefont {Peter~K.}\
  \bibnamefont {Relich}}, \bibinfo {author} {\bibfnamefont {Mark~J.}\
  \bibnamefont {Olah}}, \bibinfo {author} {\bibfnamefont {Patrick~J.}\
  \bibnamefont {Cutler}}, \ and\ \bibinfo {author} {\bibfnamefont {Keith~A.}\
  \bibnamefont {Lidke}},\ }\bibfield  {title} {\enquote {\bibinfo {title}
  {Estimation of the diffusion constant from intermittent trajectories with
  variable position uncertainties},}\ }\href {\doibase
  10.1103/PhysRevE.93.042401} {\bibfield  {journal} {\bibinfo  {journal} {Phys.
  Rev. E}\ }\textbf {\bibinfo {volume} {93}},\ \bibinfo {pages} {042401}
  (\bibinfo {year} {2016})}\BibitemShut {NoStop}%
\bibitem [{\citenamefont {Calderon}(2016)}]{calderon2016}%
  \BibitemOpen
  \bibfield  {author} {\bibinfo {author} {\bibfnamefont {Christopher~P.}\
  \bibnamefont {Calderon}},\ }\bibfield  {title} {\enquote {\bibinfo {title}
  {Motion blur filtering: A statistical approach for extracting confinement
  forces and diffusivity from a single blurred trajectory},}\ }\href {\doibase
  10.1103/PhysRevE.93.053303} {\bibfield  {journal} {\bibinfo  {journal} {Phys.
  Rev. E}\ }\textbf {\bibinfo {volume} {93}},\ \bibinfo {pages} {053303}
  (\bibinfo {year} {2016})}\BibitemShut {NoStop}%
\bibitem [{\citenamefont {Bernstein}\ and\ \citenamefont
  {Fricks}(2016)}]{bernstein2016}%
  \BibitemOpen
  \bibfield  {author} {\bibinfo {author} {\bibfnamefont {Jason}\ \bibnamefont
  {Bernstein}}\ and\ \bibinfo {author} {\bibfnamefont {John}\ \bibnamefont
  {Fricks}},\ }\bibfield  {title} {\enquote {\bibinfo {title} {Analysis of
  single particle diffusion with transient binding using particle filtering},}\
  }\href {\doibase 10.1016/j.jtbi.2016.04.013} {\bibfield  {journal} {\bibinfo
  {journal} {J. Theor. Biol.}\ } (\bibinfo {year} {2016}),\
  10.1016/j.jtbi.2016.04.013}\BibitemShut {NoStop}%
\bibitem [{\citenamefont {Koo}\ \emph {et~al.}(2015)\citenamefont {Koo},
  \citenamefont {Weitzman}, \citenamefont {Sabanaygam}, \citenamefont {van
  Golen},\ and\ \citenamefont {Mochrie}}]{koo2015}%
  \BibitemOpen
  \bibfield  {author} {\bibinfo {author} {\bibfnamefont {Peter~K.}\
  \bibnamefont {Koo}}, \bibinfo {author} {\bibfnamefont {Matthew}\ \bibnamefont
  {Weitzman}}, \bibinfo {author} {\bibfnamefont {Chandran~R.}\ \bibnamefont
  {Sabanaygam}}, \bibinfo {author} {\bibfnamefont {Kenneth~L.}\ \bibnamefont
  {van Golen}}, \ and\ \bibinfo {author} {\bibfnamefont {Simon G.~J.}\
  \bibnamefont {Mochrie}},\ }\bibfield  {title} {\enquote {\bibinfo {title}
  {Extracting diffusive states of {Rho} {GTPase} in live cells: Towards in vivo
  biochemistry},}\ }\href {\doibase 10.1371/journal.pcbi.1004297} {\bibfield
  {journal} {\bibinfo  {journal} {PLOS Comput. Biol.}\ }\textbf {\bibinfo
  {volume} {11}},\ \bibinfo {pages} {e1004297} (\bibinfo {year}
  {2015})}\BibitemShut {NoStop}%
\bibitem [{\citenamefont {Slator}\ \emph {et~al.}(2015)\citenamefont {Slator},
  \citenamefont {Cairo},\ and\ \citenamefont {Burroughs}}]{slator2015}%
  \BibitemOpen
  \bibfield  {author} {\bibinfo {author} {\bibfnamefont {Paddy~J.}\
  \bibnamefont {Slator}}, \bibinfo {author} {\bibfnamefont {Christopher~W.}\
  \bibnamefont {Cairo}}, \ and\ \bibinfo {author} {\bibfnamefont {Nigel~J.}\
  \bibnamefont {Burroughs}},\ }\bibfield  {title} {\enquote {\bibinfo {title}
  {Detection of diffusion heterogeneity in single particle tracking
  trajectories using a hidden {Markov} model with measurement noise
  propagation},}\ }\href {\doibase 10.1371/journal.pone.0140759} {\bibfield
  {journal} {\bibinfo  {journal} {PLOS ONE}\ }\textbf {\bibinfo {volume}
  {10}},\ \bibinfo {pages} {e0140759} (\bibinfo {year} {2015})}\BibitemShut
  {NoStop}%
\bibitem [{\citenamefont {Calderon}(2014)}]{calderon2014}%
  \BibitemOpen
  \bibfield  {author} {\bibinfo {author} {\bibfnamefont {Christopher~P.}\
  \bibnamefont {Calderon}},\ }\bibfield  {title} {\enquote {\bibinfo {title}
  {Data-driven techniques for detecting dynamical state changes in noisily
  measured {3D} single-molecule trajectories},}\ }\href {\doibase
  10.3390/molecules191118381} {\bibfield  {journal} {\bibinfo  {journal}
  {Molecules}\ }\textbf {\bibinfo {volume} {19}},\ \bibinfo {pages}
  {18381--18398} (\bibinfo {year} {2014})}\BibitemShut {NoStop}%
\bibitem [{\citenamefont {Chao}\ \emph {et~al.}(2013)\citenamefont {Chao},
  \citenamefont {Ram}, \citenamefont {Ward},\ and\ \citenamefont
  {Ober}}]{chao2013}%
  \BibitemOpen
  \bibfield  {author} {\bibinfo {author} {\bibfnamefont {Jerry}\ \bibnamefont
  {Chao}}, \bibinfo {author} {\bibfnamefont {Sripad}\ \bibnamefont {Ram}},
  \bibinfo {author} {\bibfnamefont {E.~Sally}\ \bibnamefont {Ward}}, \ and\
  \bibinfo {author} {\bibfnamefont {Raimund~J.}\ \bibnamefont {Ober}},\
  }\bibfield  {title} {\enquote {\bibinfo {title} {Two approximations for the
  geometric model of signal amplification in an electron-multiplying
  charge-coupled device detector},}\ }\href {\doibase 10.1117/12.2004621}
  {\bibfield  {journal} {\bibinfo  {journal} {Proc. SPIE}\ }\textbf {\bibinfo
  {volume} {8589}},\ \bibinfo {pages} {858905} (\bibinfo {year}
  {2013})}\BibitemShut {NoStop}%
\bibitem [{\citenamefont {{MacKay}}(2003)}]{mackay2003}%
  \BibitemOpen
  \bibfield  {author} {\bibinfo {author} {\bibfnamefont {David}\ \bibnamefont
  {{MacKay}}},\ }\href@noop {} {\emph {\bibinfo {title} {Information theory,
  inference, and learning algorithms}}}\ (\bibinfo  {publisher} {Cambridge
  University Press},\ \bibinfo {year} {2003})\BibitemShut {NoStop}%
\bibitem [{\citenamefont {Kirshner}\ \emph {et~al.}(2013)\citenamefont
  {Kirshner}, \citenamefont {Aguet}, \citenamefont {Sage},\ and\ \citenamefont
  {UNSER}}]{kirshner2013}%
  \BibitemOpen
  \bibfield  {author} {\bibinfo {author} {\bibfnamefont {H.}~\bibnamefont
  {Kirshner}}, \bibinfo {author} {\bibfnamefont {F.}~\bibnamefont {Aguet}},
  \bibinfo {author} {\bibfnamefont {D.}~\bibnamefont {Sage}}, \ and\ \bibinfo
  {author} {\bibfnamefont {M.}~\bibnamefont {UNSER}},\ }\bibfield  {title}
  {\enquote {\bibinfo {title} {{3-D PSF} fitting for fluorescence microscopy:
  implementation and localization application},}\ }\href {\doibase
  10.1111/j.1365-2818.2012.03675.x} {\bibfield  {journal} {\bibinfo  {journal}
  {J. Microsc.-Oxford}\ }\textbf {\bibinfo {volume} {249}},\ \bibinfo {pages}
  {13--25} (\bibinfo {year} {2013})}\BibitemShut {NoStop}%
\bibitem [{\citenamefont {Deschout}\ \emph {et~al.}(2012)\citenamefont
  {Deschout}, \citenamefont {Neyts},\ and\ \citenamefont
  {Braeckmans}}]{deschout2012}%
  \BibitemOpen
  \bibfield  {author} {\bibinfo {author} {\bibfnamefont {Hendrik}\ \bibnamefont
  {Deschout}}, \bibinfo {author} {\bibfnamefont {Kristiaan}\ \bibnamefont
  {Neyts}}, \ and\ \bibinfo {author} {\bibfnamefont {Kevin}\ \bibnamefont
  {Braeckmans}},\ }\bibfield  {title} {\enquote {\bibinfo {title} {The
  influence of movement on the localization precision of sub-resolution
  particles in fluorescence microscopy},}\ }\href {\doibase
  10.1002/jbio.201100078} {\bibfield  {journal} {\bibinfo  {journal} {J.
  Biophotonics}\ }\textbf {\bibinfo {volume} {5}},\ \bibinfo {pages} {97–109}
  (\bibinfo {year} {2012})}\BibitemShut {NoStop}%
\bibitem [{\citenamefont {Azzalini}(1985)}]{azzalini1985}%
  \BibitemOpen
  \bibfield  {author} {\bibinfo {author} {\bibfnamefont {A.}~\bibnamefont
  {Azzalini}},\ }\bibfield  {title} {\enquote {\bibinfo {title} {A class of
  distributions which includes the normal ones},}\ }\href@noop {} {\bibfield
  {journal} {\bibinfo  {journal} {Scand. J. Stat.}\ }\textbf {\bibinfo {volume}
  {12}},\ \bibinfo {pages} {171--178} (\bibinfo {year} {1985})}\BibitemShut
  {NoStop}%
\bibitem [{\citenamefont {Berglund}(2010)}]{berglund2010}%
  \BibitemOpen
  \bibfield  {author} {\bibinfo {author} {\bibfnamefont {Andrew~J.}\
  \bibnamefont {Berglund}},\ }\bibfield  {title} {\enquote {\bibinfo {title}
  {Statistics of camera-based single-particle tracking},}\ }\href {\doibase
  10.1103/PhysRevE.82.011917} {\bibfield  {journal} {\bibinfo  {journal} {Phys.
  Rev. E}\ }\textbf {\bibinfo {volume} {82}},\ \bibinfo {pages} {011917}
  (\bibinfo {year} {2010})}\BibitemShut {NoStop}%
\bibitem [{\citenamefont {Ashley}\ and\ \citenamefont
  {Andersson}(2015)}]{ashley2015}%
  \BibitemOpen
  \bibfield  {author} {\bibinfo {author} {\bibfnamefont {Trevor~T.}\
  \bibnamefont {Ashley}}\ and\ \bibinfo {author} {\bibfnamefont {Sean~B.}\
  \bibnamefont {Andersson}},\ }\bibfield  {title} {\enquote {\bibinfo {title}
  {Method for simultaneous localization and parameter estimation in particle
  tracking experiments},}\ }\href {\doibase 10.1103/PhysRevE.92.052707}
  {\bibfield  {journal} {\bibinfo  {journal} {Phys. Rev. E}\ }\textbf {\bibinfo
  {volume} {92}},\ \bibinfo {pages} {052707} (\bibinfo {year}
  {2015})}\BibitemShut {NoStop}%
\bibitem [{\citenamefont {Persson}\ \emph {et~al.}(2013)\citenamefont
  {Persson}, \citenamefont {Lind\'n}, \citenamefont {Unoson},\ and\
  \citenamefont {Elf}}]{persson2013}%
  \BibitemOpen
  \bibfield  {author} {\bibinfo {author} {\bibfnamefont {Fredrik}\ \bibnamefont
  {Persson}}, \bibinfo {author} {\bibfnamefont {Martin}\ \bibnamefont
  {Lind\'n}}, \bibinfo {author} {\bibfnamefont {Cecilia}\ \bibnamefont
  {Unoson}}, \ and\ \bibinfo {author} {\bibfnamefont {Johan}\ \bibnamefont
  {Elf}},\ }\bibfield  {title} {\enquote {\bibinfo {title} {Extracting
  intracellular diffusive states and transition rates from single-molecule
  tracking data},}\ }\href {\doibase 10.1038/nmeth.2367} {\bibfield  {journal}
  {\bibinfo  {journal} {Nat. Meth.}\ }\textbf {\bibinfo {volume} {10}},\
  \bibinfo {pages} {265--269} (\bibinfo {year} {2013})}\BibitemShut {NoStop}%
\bibitem [{\citenamefont {Liu}\ \emph {et~al.}(2013)\citenamefont {Liu},
  \citenamefont {Kromann}, \citenamefont {Krueger}, \citenamefont
  {Bewersdorf},\ and\ \citenamefont {Lidke}}]{liu2013}%
  \BibitemOpen
  \bibfield  {author} {\bibinfo {author} {\bibfnamefont {Sheng}\ \bibnamefont
  {Liu}}, \bibinfo {author} {\bibfnamefont {Emil~B.}\ \bibnamefont {Kromann}},
  \bibinfo {author} {\bibfnamefont {Wesley~D.}\ \bibnamefont {Krueger}},
  \bibinfo {author} {\bibfnamefont {Joerg}\ \bibnamefont {Bewersdorf}}, \ and\
  \bibinfo {author} {\bibfnamefont {Keith~A.}\ \bibnamefont {Lidke}},\
  }\bibfield  {title} {\enquote {\bibinfo {title} {Three dimensional single
  molecule localization using a phase retrieved pupil function},}\ }\href
  {\doibase 10.1364/OE.21.029462} {\bibfield  {journal} {\bibinfo  {journal}
  {Opt. Express}\ }\textbf {\bibinfo {volume} {21}},\ \bibinfo {pages} {29462}
  (\bibinfo {year} {2013})}\BibitemShut {NoStop}%
\bibitem [{\citenamefont {Pavani}\ \emph {et~al.}(2009)\citenamefont {Pavani},
  \citenamefont {Thompson}, \citenamefont {Biteen}, \citenamefont {Lord},
  \citenamefont {Liu}, \citenamefont {Twieg}, \citenamefont {Piestun},\ and\
  \citenamefont {Moerner}}]{pavani2009}%
  \BibitemOpen
  \bibfield  {author} {\bibinfo {author} {\bibfnamefont {S.~R.~P.}\
  \bibnamefont {Pavani}}, \bibinfo {author} {\bibfnamefont {M.~A.}\
  \bibnamefont {Thompson}}, \bibinfo {author} {\bibfnamefont {J.~S.}\
  \bibnamefont {Biteen}}, \bibinfo {author} {\bibfnamefont {S.~J.}\
  \bibnamefont {Lord}}, \bibinfo {author} {\bibfnamefont {N.}~\bibnamefont
  {Liu}}, \bibinfo {author} {\bibfnamefont {R.~J.}\ \bibnamefont {Twieg}},
  \bibinfo {author} {\bibfnamefont {R.}~\bibnamefont {Piestun}}, \ and\
  \bibinfo {author} {\bibfnamefont {W.~E.}\ \bibnamefont {Moerner}},\
  }\bibfield  {title} {\enquote {\bibinfo {title} {Three-dimensional,
  single-molecule fluorescence imaging beyond the diffraction limit by using a
  double-helix point spread function},}\ }\href {\doibase
  10.1073/pnas.0900245106} {\bibfield  {journal} {\bibinfo  {journal} {Proc.
  Natl. Acad. Sci. U.S.A.}\ }\textbf {\bibinfo {volume} {106}},\ \bibinfo
  {pages} {2995--2999} (\bibinfo {year} {2009})}\BibitemShut {NoStop}%
\bibitem [{\citenamefont {Chow}(2009)}]{chow2009}%
  \BibitemOpen
  \bibfield  {author} {\bibinfo {author} {\bibfnamefont {Winston~C.}\
  \bibnamefont {Chow}},\ }\bibfield  {title} {\enquote {\bibinfo {title}
  {Brownian bridge},}\ }\href {\doibase 10.1002/wics.38} {\bibfield  {journal}
  {\bibinfo  {journal} {WIREs Comp. Stat.}\ }\textbf {\bibinfo {volume} {1}},\
  \bibinfo {pages} {325–332} (\bibinfo {year} {2009})}\BibitemShut {NoStop}%
\bibitem [{\citenamefont {Ghahramani}\ and\ \citenamefont
  {Jordan}(1997)}]{ghahramani1997}%
  \BibitemOpen
  \bibfield  {author} {\bibinfo {author} {\bibfnamefont {Zoubin}\ \bibnamefont
  {Ghahramani}}\ and\ \bibinfo {author} {\bibfnamefont {Michael~I.}\
  \bibnamefont {Jordan}},\ }\bibfield  {title} {\enquote {\bibinfo {title}
  {Factorial hidden {Markov} models},}\ }\href {\doibase
  10.1023/A:1007425814087} {\bibfield  {journal} {\bibinfo  {journal} {Mach.
  Learn.}\ }\textbf {\bibinfo {volume} {29}},\ \bibinfo {pages} {245--273}
  (\bibinfo {year} {1997})}\BibitemShut {NoStop}%
\bibitem [{\citenamefont {Bishop}(2006)}]{bishop2006}%
  \BibitemOpen
  \bibfield  {author} {\bibinfo {author} {\bibfnamefont {Christopher}\
  \bibnamefont {Bishop}},\ }\href@noop {} {\emph {\bibinfo {title} {Pattern
  recognition and machine learning}}}\ (\bibinfo  {publisher} {Springer},\
  \bibinfo {address} {New York},\ \bibinfo {year} {2006})\BibitemShut {NoStop}%
\bibitem [{\citenamefont {Rabiner}\ and\ \citenamefont
  {Juang}(1986)}]{rabiner1986}%
  \BibitemOpen
  \bibfield  {author} {\bibinfo {author} {\bibfnamefont {L.~R.}\ \bibnamefont
  {Rabiner}}\ and\ \bibinfo {author} {\bibfnamefont {B.~H.}\ \bibnamefont
  {Juang}},\ }\bibfield  {title} {\enquote {\bibinfo {title} {An introduction
  to hidden {Markov} models},}\ }\href@noop {} {\bibfield  {journal} {\bibinfo
  {journal} {IEEE ASSp Magazine}\ ,\ \bibinfo {pages} {4--16}} (\bibinfo {year}
  {1986})}\BibitemShut {NoStop}%
\bibitem [{\citenamefont {Bronson}\ \emph {et~al.}(2009)\citenamefont
  {Bronson}, \citenamefont {Fei}, \citenamefont {Hofman}, \citenamefont
  {Gonzalez~Jr.},\ and\ \citenamefont {Wiggins}}]{bronson2009}%
  \BibitemOpen
  \bibfield  {author} {\bibinfo {author} {\bibfnamefont {Jonathan~E.}\
  \bibnamefont {Bronson}}, \bibinfo {author} {\bibfnamefont {Jingyi}\
  \bibnamefont {Fei}}, \bibinfo {author} {\bibfnamefont {Jake~M.}\ \bibnamefont
  {Hofman}}, \bibinfo {author} {\bibfnamefont {Ruben~L.}\ \bibnamefont
  {Gonzalez~Jr.}}, \ and\ \bibinfo {author} {\bibfnamefont {Chris~H.}\
  \bibnamefont {Wiggins}},\ }\bibfield  {title} {\enquote {\bibinfo {title}
  {Learning rates and states from biophysical time series: A {Bayesian}
  approach to model selection and single-molecule {FRET} data},}\ }\href
  {\doibase 10.1016/j.bpj.2009.09.031} {\bibfield  {journal} {\bibinfo
  {journal} {Biophys. J.}\ }\textbf {\bibinfo {volume} {97}},\ \bibinfo {pages}
  {3196--3205} (\bibinfo {year} {2009})}\BibitemShut {NoStop}%
\bibitem [{\citenamefont {Johnson}\ \emph {et~al.}(2014)\citenamefont
  {Johnson}, \citenamefont {Meent}, \citenamefont {Phillips}, \citenamefont
  {Wiggins},\ and\ \citenamefont {Lind\'en}}]{johnson2014}%
  \BibitemOpen
  \bibfield  {author} {\bibinfo {author} {\bibfnamefont {Stephanie}\
  \bibnamefont {Johnson}}, \bibinfo {author} {\bibfnamefont {Jan-Willem
  van~de}\ \bibnamefont {Meent}}, \bibinfo {author} {\bibfnamefont {Rob}\
  \bibnamefont {Phillips}}, \bibinfo {author} {\bibfnamefont {Chris~H.}\
  \bibnamefont {Wiggins}}, \ and\ \bibinfo {author} {\bibfnamefont {Martin}\
  \bibnamefont {Lind\'en}},\ }\bibfield  {title} {\enquote {\bibinfo {title}
  {Multiple {LacI}-mediated loops revealed by {Bayesian} statistics and
  tethered particle motion},}\ }\href {\doibase 10.1093/nar/gku563} {\bibfield
  {journal} {\bibinfo  {journal} {Nucl. Acids Res.}\ }\textbf {\bibinfo
  {volume} {42}},\ \bibinfo {pages} {10265--10277} (\bibinfo {year}
  {2014})}\BibitemShut {NoStop}%
\bibitem [{\citenamefont {Burnham}\ and\ \citenamefont
  {Anderson}(2013)}]{burnham}%
  \BibitemOpen
  \bibfield  {author} {\bibinfo {author} {\bibfnamefont {Kenneth~P.}\
  \bibnamefont {Burnham}}\ and\ \bibinfo {author} {\bibfnamefont {David~R.}\
  \bibnamefont {Anderson}},\ }\href@noop {} {\emph {\bibinfo {title} {Model
  Selection and Multimodel Inference: {A} Practical Information-Theoretic
  Approach}}}\ (\bibinfo  {publisher} {Springer},\ \bibinfo {year}
  {2013})\BibitemShut {NoStop}%
\bibitem [{\citenamefont {Chao}\ \emph {et~al.}(2015)\citenamefont {Chao},
  \citenamefont {Ram}, \citenamefont {Lee}, \citenamefont {Ward},\ and\
  \citenamefont {Ober}}]{chao2015}%
  \BibitemOpen
  \bibfield  {author} {\bibinfo {author} {\bibfnamefont {Jerry}\ \bibnamefont
  {Chao}}, \bibinfo {author} {\bibfnamefont {Sripad}\ \bibnamefont {Ram}},
  \bibinfo {author} {\bibfnamefont {Taiyoon}\ \bibnamefont {Lee}}, \bibinfo
  {author} {\bibfnamefont {E.~Sally}\ \bibnamefont {Ward}}, \ and\ \bibinfo
  {author} {\bibfnamefont {Raimund~J.}\ \bibnamefont {Ober}},\ }\bibfield
  {title} {\enquote {\bibinfo {title} {Investigation of the numerics of point
  spread function integration in single molecule localization},}\ }\href
  {\doibase 10.1364/OE.23.016866} {\bibfield  {journal} {\bibinfo  {journal}
  {Opt. Expr.}\ }\textbf {\bibinfo {volume} {23}},\ \bibinfo {pages} {16866}
  (\bibinfo {year} {2015})}\BibitemShut {NoStop}%
\bibitem [{\citenamefont {Cram{\'e}r}(1945)}]{cramer1945}%
  \BibitemOpen
  \bibfield  {author} {\bibinfo {author} {\bibfnamefont {Harald}\ \bibnamefont
  {Cram{\'e}r}},\ }\href@noop {} {\emph {\bibinfo {title} {Mathematical methods
  of statistics}}},\ Vol.~\bibinfo {volume} {9}\ (\bibinfo  {publisher}
  {Princeton university press},\ \bibinfo {year} {1945})\BibitemShut {NoStop}%
\bibitem [{\citenamefont {Rao}(1945)}]{rao1945}%
  \BibitemOpen
  \bibfield  {author} {\bibinfo {author} {\bibfnamefont {C~Radhakrishna}\
  \bibnamefont {Rao}},\ }\bibfield  {title} {\enquote {\bibinfo {title}
  {Information and accuracy attainable in the estimation of statistical
  parameters},}\ }\href@noop {} {\bibfield  {journal} {\bibinfo  {journal}
  {Bull Calcutta. Math. Soc.}\ }\textbf {\bibinfo {volume} {37}},\ \bibinfo
  {pages} {81--91} (\bibinfo {year} {1945})}\BibitemShut {NoStop}%
\bibitem [{\citenamefont {MacKay}(1998)}]{mackay1998}%
  \BibitemOpen
  \bibfield  {author} {\bibinfo {author} {\bibfnamefont {David J.~C.}\
  \bibnamefont {MacKay}},\ }\bibfield  {title} {\enquote {\bibinfo {title}
  {Choice of basis for laplace approximation},}\ }\href {\doibase
  10.1023/A:1007558615313} {\bibfield  {journal} {\bibinfo  {journal} {Mach.
  Learn.}\ }\textbf {\bibinfo {volume} {33}},\ \bibinfo {pages} {77--86}
  (\bibinfo {year} {1998})}\BibitemShut {NoStop}%
\end{thebibliography}
